\pdfoutput=1

\newif\ifdraft\draftfalse   
\newif\ifanon\anonfalse     
\newif\ifcamera\cameratrue  
\newif\ifsooner\soonerfalse 
\newif\iflater\laterfalse   
\newif\iffull\fullfalse     
\newif\ifccs\ccstrue        
\newif\ifae\aefalse         
\newif\ifneedspace\needspacetrue
\newif\ifallcites\allcitesfalse 
\newif\ifbackref\backreffalse
\makeatletter \@input{texdirectives} \makeatother

\ifccs 
  \ifcamera
    \documentclass[sigconf]{acmart}
    \AtBeginDocument{%
    }
  \else
    \ifanon
      \documentclass[sigconf,screen,nonacm,review,anonymous]{acmart}
    \else
      \ifae
      \documentclass[sigconf,screen,nonacm,review]{acmart}
      \else
      \documentclass[sigconf,screen,nonacm]{acmart}
      \fi
    \fi
  \fi
\else 
  \ifcamera
    \documentclass[acmsmall,screen,nonacm]{acmart}
  \else
    \ifanon
      \documentclass[acmsmall,screen,nonacm,review,anonymous]{acmart}
    \else
      \documentclass[acmsmall,screen,nonacm]{acmart}
    \fi
  \fi
\fi

\usepackage{booktabs}   
\usepackage{subcaption} 
\usepackage{wrapfig}

\usepackage[utf8]{inputenc}
\usepackage{xspace}
\usepackage{combelow}

\usepackage{mathpartir}

\usepackage[normalem]{ulem}

\usepackage{bm} 

\usepackage{titlesec}
\titlespacing*{\section}{0pt}{1.5ex plus 0ex minus .2ex}{0.3ex plus 0ex minus 0.2ex}
\titlespacing*{\subsection}{0pt}{1.5ex plus 0ex minus .2ex}{0.3ex plus 0ex minus 0.2ex}

\renewcommand{\paragraph}[1]{\ifneedspace\else\smallskip\fi{\bf #1.}\;}

\newcommand{\citeFull}[2]{\ifallcites\cite{#1,#2}\else\cite{#1}\fi}

\usepackage{tikz}
\usetikzlibrary{matrix,arrows,decorations.pathmorphing,positioning}

\tikzset{
    to*/.style={
        shorten >=.25em,#1-to,
        to path={-- node[inner sep=0pt,at end,sloped] {${}^*$} (\tikztotarget) \tikztonodes}
    },
    to*/.default=
}

\makeatletter
\pgfarrowsdeclare{to*}{to*}
{
  \pgfutil@tempdima=-0.84pt%
  \advance\pgfutil@tempdima by-1.3\pgflinewidth%
  \pgfutil@tempdimb=0.21pt%
  \advance\pgfutil@tempdimb by.625\pgflinewidth%
  \advance\pgfutil@tempdimb by2.5pt%
  \pgfarrowsleftextend{+\pgfutil@tempdima}
  \pgfarrowsrightextend{+\pgfutil@tempdimb}
}
{
  \pgfutil@tempdima=0.28pt%
  \advance\pgfutil@tempdima by.3\pgflinewidth%
  \pgfsetlinewidth{0.8\pgflinewidth}
  \pgfsetdash{}{+0pt}
  \pgfsetroundcap
  \pgfsetroundjoin
  \pgfpathmoveto{\pgfqpoint{-3\pgfutil@tempdima}{4\pgfutil@tempdima}}
  \pgfpathcurveto
  {\pgfqpoint{-2.75\pgfutil@tempdima}{2.5\pgfutil@tempdima}}
  {\pgfqpoint{0pt}{0.25\pgfutil@tempdima}}
  {\pgfqpoint{0.75\pgfutil@tempdima}{0pt}}
  \pgfpathcurveto
  {\pgfqpoint{0pt}{-0.25\pgfutil@tempdima}}
  {\pgfqpoint{-2.75\pgfutil@tempdima}{-2.5\pgfutil@tempdima}}
  {\pgfqpoint{-3\pgfutil@tempdima}{-4\pgfutil@tempdima}}
  \pgfusepathqstroke
  \begingroup
    \pgftransformxshift{2.5pt}
    \pgftransformyshift{2pt}
    \pgftransformscale{.7}
    \pgfuseplotmark{asterisk}
  \endgroup
}

\pgfarrowsdeclare{*to}{*to}
{
  \pgfutil@tempdima=-0.84pt%
  \advance\pgfutil@tempdima by-1.3\pgflinewidth%
  \pgfutil@tempdimb=0.21pt%
  \advance\pgfutil@tempdimb by.625\pgflinewidth%
  \advance\pgfutil@tempdimb by2.5pt%
  \pgfarrowsleftextend{+\pgfutil@tempdima}
  \pgfarrowsrightextend{+\pgfutil@tempdimb}
}
{
  \pgfutil@tempdima=0.28pt%
  \advance\pgfutil@tempdima by.3\pgflinewidth%
  \pgfsetlinewidth{0.8\pgflinewidth}
  \pgfsetdash{}{+0pt}
  \pgfsetroundcap
  \pgfsetroundjoin
  \pgfpathmoveto{\pgfqpoint{-3\pgfutil@tempdima}{4\pgfutil@tempdima}}
  \pgfpathcurveto
  {\pgfqpoint{-2.75\pgfutil@tempdima}{2.5\pgfutil@tempdima}}
  {\pgfqpoint{0pt}{0.25\pgfutil@tempdima}}
  {\pgfqpoint{0.75\pgfutil@tempdima}{0pt}}
  \pgfpathcurveto
  {\pgfqpoint{0pt}{-0.25\pgfutil@tempdima}}
  {\pgfqpoint{-2.75\pgfutil@tempdima}{-2.5\pgfutil@tempdima}}
  {\pgfqpoint{-3\pgfutil@tempdima}{-4\pgfutil@tempdima}}
  \pgfusepathqstroke
  \begingroup
    \pgftransformxshift{2.5pt}
    \pgftransformyshift{-2pt}
    \pgftransformscale{.7}
    \pgfuseplotmark{asterisk}
  \endgroup
}

\makeatother

\usepackage[inline]{enumitem}
\newlist{inlist}{enumerate*}{1}
\setlist[inlist]{label=(\arabic*)}

\usepackage{relsize} 

\usepackage{proof}

\newcommand*{\EG}{e.g.,\xspace}
\newcommand*{\IE}{i.e.,\xspace}
\newcommand*{\ETAL}{{\em et al.}}
\newcommand*{\ETC}{etc.\xspace}

\newcommand\fstar{F$^{\star}$\xspace}

\def\Snospace~{\S{}}

\ifneedspace
\makeatletter
\def\thm@space@setup{\thm@preskip=0pt
\thm@postskip=0pt}
\makeatother

\newtheoremstyle{newstyle}
{1pt} 
{0pt} 
{\mdseries} 
{} 
{\bfseries} 
{.} 
{ } 
{} 

\theoremstyle{newstyle}
\fi

\newtheorem{assumption}{Assumption}
\newtheorem{definition}{Definition}
\newtheorem{axiom}{Axiom}

\newcommand{\ii}[1]{\textit{#1}}

\newcommand{\rscdcmd}[0]{\ii{RSC}\ensuremath{^\ii{DC}_\ii{MD}}\xspace}

\newcommand{\cmp}[1]{#1\hspace{-0.35em}\downarrow}
\newcommand{\back}[1]{#1\hspace{-0.35em}\uparrow}

\newcommand{\clight}{Clight\xspace}
\newcommand{\RISCV}{RISC\nobreakdash-V\xspace}

\newcommand{\src}[1]{\mathbf{\color{RoyalBlue}#1}}
\newcommand{\trg}[1]{\mathsf{\color{BrickRed}#1}}
\newcommand{\linksrc}{\mathbin{\color{RoyalBlue}\boldsymbol{\bowtie}}}
\newcommand{\biglinksrc}{\mathlarger{\mathlarger{\linksrc}}}
\newcommand{\linktrg}{\mathbin{\color{BrickRed}\bowtie}}
\newcommand{\semsrc}{\mathrel{\mathrel{\color{RoyalBlue}\boldsymbol{\rightsquigarrow}}}}
\newcommand{\semtrg}{\mathrel{\mathrel{\color{BrickRed}\rightsquigarrow}}}

\newcommand{\undefi}{\mathsf{Undef}}

\definecolor{dkblue}{rgb}{0,0,.6}

\newcommand{\comm}[3]{\ifdraft{\color{#1}[#2: #3]}\fi}

\newcommand{\jt}[1]{\comm{violet}{JT}{#1}}
\newcommand{\rb}[1]{\comm{orange}{RB}{#1}}

\newcommand{\ch}[1]{\comm{teal}{CH}{#1}}

\newcommand{\apt}[1]{\comm{purple}{APT}{#1}}

\ifcamera
  \copyrightyear{2024}
  \acmYear{2024}
  \setcopyright{rightsretained}
  \acmConference[CCS '24]{Proceedings of the 2024 ACM SIGSAC Conference on Computer and Communications Security}{October 14--18, 2024}{Salt Lake City, UT, USA}
  \acmBooktitle{Proceedings of the 2024 ACM SIGSAC Conference on Computer and Communications Security (CCS '24), October 14--18, 2024, Salt Lake City, UT, USA}
  \acmDOI{10.1145/3658644.3670288}
  \acmISBN{979-8-4007-0636-3/24/10}
\else
  \setcopyright{none}
\fi

\ifcamera
\makeatletter
\gdef\@copyrightpermission{
 \begin{minipage}{0.3\columnwidth}
  \href{https://creativecommons.org/licenses/by/4.0/}{\includegraphics[width=0.90\textwidth]{4ACM-CC-by-88x31.pdf}}
 \end{minipage}\hfill
 \begin{minipage}{0.7\columnwidth}
  \href{https://creativecommons.org/licenses/by/4.0/}{This work is licensed under a Creative Commons Attribution International 4.0 License.}
 \end{minipage}
 \vspace{5pt}
}
\makeatother
\fi

\ifcamera
\else
\fi



\begin{document}

\title[SECOMP: Formally Secure Compilation of Compartmentalized C Programs]{SECOMP: Formally Secure Compilation\\of Compartmentalized C Programs}


\ifanon\else
\ifcamera 
\author{J\'er\'emy Thibault}
  \affiliation{\institution{Max Planck Institute for\\Security and Privacy (MPI-SP)}\city{Bochum}\country{Germany}}
  \email{jeremy.thibault@mpi-sp.org}
  \orcid{0009-0008-5112-3269}
\author{Roberto Blanco}
  \affiliation{\institution{Max Planck Institute for\\Security and Privacy (MPI-SP)}\city{Bochum}\country{Germany}}
  \email{roberto.blanco@mpi-sp.org}
  \orcid{0000-0002-1017-0391}
\author{Dongjae Lee}
  \affiliation{\institution{Seoul National University}\ifcamera\city{Seoul}\fi\country{South Korea}}
  \email{dongjae.lee@sf.snu.ac.kr}
  \orcid{0000-0003-2576-1220}
\author{Sven Argo}
  \affiliation{\institution{Ruhr University Bochum}\ifcamera\city{Bochum}\fi\country{Germany}}
  \email{sven.argo@rub.de}
  \orcid{0009-0009-5544-6417}
\author{Arthur Azevedo de Amorim}
  \affiliation{\institution{Rochester Institute of Technology}\ifcamera\city{Rochester, NY}\fi\country{USA}}
  \email{arthur.aa@gmail.com}
  \orcid{000-0001-9916-6614}
\author{Aïna Linn Georges}
  \affiliation{\institution{Max Planck Institute for\\Software Systems (MPI-SWS)}\city{Saarbrücken}\country{Germany}}
  \email{algeorges@mpi-sws.org}
  \orcid{0000-0002-5951-4642}
\author{C\u{a}t\u{a}lin Hri\cb{t}cu}
  \affiliation{\institution{Max Planck Institute for\\Security and Privacy (MPI-SP)}\city{Bochum}\country{Germany}}
  \email{catalin.hritcu@mpi-sp.org}
  \orcid{0000-0001-8919-8081}
\author{Andrew Tolmach}
  \affiliation{\institution{Portland State University}\ifcamera\city{Portland, OR}\fi\country{USA}}
  \email{tolmach@pdx.edu}
  \orcid{0000-0002-0748-2044}
\else
\author{J\'er\'emy Thibault$^1$ \quad
        Roberto Blanco$^1$ \quad
        Dongjae Lee$^{2}$ \quad
        Sven Argo$^3$ \\[0.3em]
        Arthur Azevedo de Amorim$^4$ \quad
        Aïna Linn Georges$^5$ \quad
        C\u{a}t\u{a}lin Hri\cb{t}cu$^1$ \quad
        Andrew Tolmach$^6$}
\affiliation{\vspace{0.8em}$^1$MPI-SP, Bochum, Germany \qquad
             $^2$Seoul National University, South Korea \qquad
             $^3$Ruhr University Bochum, Germany \\[0.5em]
             $^4$Rochester Institute of Technology, USA \qquad
             $^5$MPI-SWS, Saarbrücken, Germany \qquad
             $^6$Portland State University, USA\vspace{1em}
  \country{} 
}
\fi
\fi

\ifanon\else
\ifcamera\renewcommand{\shortauthors}{Jérémy Thibault et al.}
\else\renewcommand{\shortauthors}{Jérémy Thibault, Roberto Blanco, Dongjae Lee, et al.}
\fi\fi

\begin{abstract}
Undefined behavior in C often causes devastating security vulnerabilities. One practical mitigation is compartmentalization, which allows developers to structure large programs into mutually distrustful compartments with clearly specified privileges and interactions. In this paper we introduce SECOMP, a compiler for compartmentalized C code that comes with machine-checked proofs guaranteeing that the scope of undefined behavior is restricted to the compartments that encounter it and become dynamically compromised. These guarantees are formalized as the preservation of safety properties against adversarial contexts, a secure compilation criterion similar to full abstraction, and this is the first time such a strong criterion is proven for a mainstream programming language. To achieve this we extend the languages of the CompCert verified C compiler with isolated compartments that can only interact via procedure calls and returns, as specified by cross-compartment interfaces. We adapt the passes and optimizations of CompCert as well as their correctness proofs to this compartment-aware setting. We then use compiler correctness as an ingredient in a larger secure compilation proof that involves several proof engineering novelties, needed to scale formally secure compilation up to a C compiler.


\end{abstract}




\maketitle


\section{Introduction}
\label{sec:intro}


Undefined behavior is endemic in the C language: buffer overflows,
use after frees, double frees, signed integer overflows,
invalid type casts, various concurrency bugs,
\ETC, cause mainstream C compilers to produce code that can behave completely arbitrarily.
This leads to devastating security vulnerabilities that are often remotely
exploitable, and both Microsoft and Chrome report that around 70\% of their high
severity security bugs are caused by undefined behavior due to memory safety violations
alone~\citeFull{memory-unsafety-microsoft-talk, memory-unsafety-chrome}{memory-unsafety-chrome-zdnet}.

A strong practical mitigation against such vulnerabilities is
{\em compartmentalization}~\citeFull{GudkaWACDLMNR15, wedge_nsdi2008, Kilpatrick03}{VasilakisKRDDS18},
which allows developers to structure large programs
into mutually distrustful compartments that have clearly specified privileges
and that can only interact via well-defined interfaces.
This way, the compromise of some compartments has a limited impact on the
security of the whole program.
This intuitive increase in security has made compartmentalization and the
compartment isolation technologies used to enforce it become widely deployed in
practice; \EG all major web browsers today use both process-level privilege
separation~\cite{Kilpatrick03, GudkaWACDLMNR15, wedge_nsdi2008} to isolate tabs
and plugins~\cite{ReisG09}, and software fault isolation (SFI)~\cite{sfi_sosp1993,
  YeeSDCMOONF10, Tan17, MorrisettTTTG12, ZhaoLSR11}
to sandbox WebAssembly modules~\citeFull{HaasRSTHGWZB17}{KolosickNJWLGJS22,
  JohnsonLZGNSSB23, JohnsonTANBLMSS21, VanHattumPFSB24}.

In this paper, we investigate how to provide strong {\em formal guarantees}
for compartmentalized C source code by making the C compiler aware of compartments.{}
We follow Abate~\ETAL~\cite{AbateABEFHLPST18}, who argue that a
compartment-aware compiler for an unsafe language can restrict the scope of
undefined behavior both (a)~spatially to just the compartments that
encounter it\ifanon\else~\cite{JuglaretHAEP16}\fi, and (b)~temporally by
still providing protection to each compartment up to the point in time when it
encounters undefined behavior. Abate~\ETAL{} formalize this intuition as a
variant of a general {\em secure compilation} criterion called {\em Robust Safety
Preservation (RSP)}~\citeFull{AbateBGHPT19, difftraces, PatrignaniG21}{PatrignaniG17}.
Their RSP variant ensures that any low-level attack against a compiled
program's safety properties mounted by compartments dynamically compromised by undefined behavior
could also have been mounted at the source level by arbitrary compartments with the same
interface and privileges, while staying in the secure fragment of the source
semantics, without undefined behavior.
This strong 
guarantee allows source-level security reasoning about compartmentalized programs that have undefined
behavior, and thus for which the C standard and the usual C compilers would
provide no guarantees whatsoever.
%

Such strong formal guarantees are, however, notoriously challenging to achieve in practice
and to prove mathematically. RSP~\citeFull{AbateBGHPT19, difftraces, PatrignaniG21}{PatrignaniG17}
belongs to the same class of secure compilation criteria as full
abstraction~\cite{Abadi99, MarcosSurvey}, for which simple and intuitive but wrong conjectures
have sometimes survived for decades~\cite{DevriesePP18}, and for which careful paper
proofs can take hundreds of pages even for very simple languages and
compilers~\cite{akram-capabileptrs, JuglaretHAEP16}.
Such proofs are generally so challenging that no compiler for a mainstream
programming language that is guaranteed to achieve any such secure compilation
criterion has ever been built.
Moreover, such secure compilation proofs are at the moment often only done on
paper~\citeFull{AgtenSJP12, JuglaretHAEP16, PatrignaniC15, PatrignaniASJCP15, 
  MarcosSurvey, AbateBGHPT19, PatrignaniG21, BusiNBGDMP21,
  akram-capabileptrs}{AbadiFG02, AbadiP12, AhmedB08, AhmedB11, FournetSCDSL13,
  NewBA16,JagadeesanPRR11, PatrignaniDP16, PatrignaniG17},
even though at the scale of a realistic compiler, paper
proofs would be impossible to trust, construct, and maintain.
All this stands in stark contrast to {\em compiler correctness}: 
CompCert~\citeFull{Leroy09b}{Leroy09}---a realistic C compiler that comes with a
machine-checked correctness proof in the Coq proof assistant---has already existed for
more than a decade and is used in practice in highly safety-critical
applications~\cite{CompCert-ERTS-2018}.

In this paper we take an important step towards bridging this gap by devising
SECOMP, a formally secure compiler for compartmentalized C code.
To do this, we extend the CompCert compiler and its correctness proof to handle
isolated compartments that interact only via procedure calls and returns.
Although compiler correctness by itself is definitely not enough to prove secure
compilation, since it
gives up on programs with undefined behavior,
we use it as one key ingredient for such a proof. 
For this we adopt the high-level proof structure proposed by
Abate~\ETAL~\cite{AbateABEFHLPST18}, who showed how proving their RSP variant
can be reduced to showing compiler correctness together with three
security-related properties: {\em back-translation},
{\em recomposition}, and {\em blame} (explained in the next section, \autoref{sec:background}).
Proving these properties at scale and achieving formally secure compilation for
a compiler for a mainstream programming language were open research challenges,
which we solve in this work by bringing the following {\bf novel contributions}:

\begin{itemize}[leftmargin=0pt,nosep,label=$\blacktriangleright$]

\item We devise the SECOMP compiler for compartmentalized C programs 
to \RISCV assembly by
extending the syntax and semantics of all the 10 languages of CompCert
with the abstraction of isolated compartments that can only
interact via procedure calls, as specified by cross-compartment interfaces.
For CompCert's \RISCV assembly we propose an enforcement-independent
characterization of C compartments that relies on a new shadow stack
to ensure the well-bracketedness of cross-compartment control flow.
%
We adapt all 19 passes and all optimizations of CompCert to this extension,
except cross-compartment inlining and tail-calls, which we 
disallow.

\item In addition to passing scalar values to each other, 
our compartments can also perform input and output (IO), which was not the case
in the very simple languages studied by Abate~\ETAL~\cite{AbateABEFHLPST18}.
Our IO model allows pointers to global buffers of scalars to be passed to the
system calls implementing IO and also allows these buffers to be changed
nondeterministically by these system calls, which goes beyond what was
previously possible in CompCert's IO model.

\item We extend CompCert's large-scale compiler correctness proof to
account for these changes so that we can use it to show secure compilation.
Our extension of the correctness proof is elegant and relatively small, even
though two of our changes to the semantics of the CompCert languages
are substantial:
(1) we extend the CompCert memory model with compartments, and
(2) we extend the CompCert trace model with events recording cross-compartment
calls and returns, as needed for the secure compilation proof.



\item We develop a 
  secure compilation proof for SECOMP in Coq,
from Clight, the first intermediate language of CompCert and
featuring a determinate~\citeFull{Engelfriet85}{milner82,ChevalCD13}
semantics (as opposed to CompCert C),
down to our extension of CompCert's \RISCV assembly.
This proof shows the RSP variant of Abate~\ETAL~\cite{AbateABEFHLPST18},
capturing the secure compilation of 
mutually distrustful C
compartments that can be dynamically compromised by undefined behavior.
We are the first to prove such a strong secure compilation criterion for
a mainstream programming language, which makes
this a milestone for secure compilation.
%

\item To scale up the secure compilation proofs to SECOMP,
we introduce several proof engineering  
novelties: 
(1)~Because SECOMP uses the memory model of CompCert~\cite{LeroyB08}
(extended with compartments),
the novel simulation invariants we devise to prove security
have to make use of the sophisticated {\em memory injections} of CompCert~\cite{LeroyB08},
which provide a fine-grained characterization of the way memory is transformed
during compilation.
(2)~
%
For back-translation, because system calls may read
global buffers, we extend traces with \emph{informative events}, which record \emph{memory
  deltas}---\IE changes to global buffers happening during silent
steps---and we use those to establish memory injections to prove
correctness of the system calls the back-translation generates.
%
(3)~For recomposition, we propose a more principled way of proving the required three-way
simulation by 
defining 8 simulation diagrams and providing a general
proof that together they imply recomposition.
Despite 
the realistic \RISCV{} instruction
set,{}
we use these diagrams to provide a 
compact proof of recomposition showing that our \RISCV{} assembly semantics
securely characterizes the compartment abstraction.
We discovered that, for recomposition to hold, the stack-spilled call arguments
spilled must be protected, so a malicious caller cannot exploit callbacks to
covertly change arguments of a previous call.
%

\item The SECOMP secure compilation proofs end at our extension of CompCert's
\RISCV assembly, which is the language where CompCert's compiler correctness
proofs also end, and whose semantics still maintains CompCert's
block-based memory model.
As mentioned above, to this language's semantics we added the extra abstraction
of isolated compartments, which formally defines {\em what} compartment
isolation enforcement should do, but which leaves the {\em how} to lower-level
enforcement mechanisms working with a more concrete view of memory as an array
of bytes~\cite{WangWS19,WangXWS20} and potentially making use of hardware
security features.
We additionally show that the compartment isolation abstraction can be enforced
at a lower level by designing and prototyping an unverified backend targeting a
variant of the CHERI capability machine~\cite{CHERI-ISA}.
For this we extend a recently proposed efficient calling convention enforcing stack
safety~\cite{GeorgesTB22} to our setting of mutually distrustful compartments
by introducing capability-protected wrappers to clear registers on
calls and returns and to prevent capabilities from being passed between
potentially compromised compartments.
Various other enforcement mechanisms should be possible though, including
SFI~\citeFull{sfi_sosp1993, YeeSDCMOONF10, Tan17}{KolosickNJWLGJS22, JohnsonLZGNSSB23} and tagged
architectures~\cite{micropolicies2015, pump_asplos2015}, as shown in a
simpler setting by Abate~\ETAL~\cite{AbateABEFHLPST18}.
At the moment all these lower-level backends are, however,
unverified, and extending the secure compilation proofs to cover them is a
formidable research challenge that we leave as future work (\autoref{sec:future}).

\end{itemize}

\paragraph{Long version} A long version of this paper with additional details is
available at {\tt \url{http://arxiv.org/abs/2401.16277}}

\paragraph{Artifact}
The SECOMP formally secure compiler
\ifanon was provided as a supplementary material with this submission.
\else is available at \url{https://github.com/secure-compilation/SECOMP}
and as a permanently archived artifact at \url{https://doi.org/10.5281/zenodo.11007679}.
\fi
%
SECOMP adds $\sim$43k LoC 
on top of CompCert, mostly in proofs, when excluding the unverified backend.
%
In more detail, our extensions to CompCert 3.12 and its correctness
comprise $\sim$7k LoC of specs and
$\sim$11k of proofs, an increase of 7.2\% and 22.3\% respectively.
%
In addition, back-translation involves $\sim$5k LoC of specs and $\sim$6k of
proofs; recomposition $\sim$1k LoC of specs and $\sim$8k of proofs; and blame
$\sim$1k LoC of specs and $\sim$4k of proofs.
%
%
%

These machine-checked proofs are generally
complete and include no further axioms beyond those already
existing in CompCert~\cite{MonniauxB22}, or small adaptions thereof to account
for the addition of compartments to the compiler.
One exception to this is an axiom assuming that CompCert can successfully
compile the results of our back-translation, which would be very tedious to
prove in general, but which we have instead thoroughly tested (\autoref{sec:compiling-bt}).
The other current gap in our Coq formalization is about connecting
compiler correctness, recomposition, back-translation, and blame into a single
mechanized secure compilation result (\autoref{thm:rscdcmd}); instead at the
moment the top-level proof and all the steps are complete,
but they are not integrated, and back-translation and blame are still on separate branches.
%
This is further documented in {\tt README.md}.

Finally, the unverified CHERI \RISCV{} backend
consists of $\sim$13k LoC of specs and $\sim$8k of proofs,
which are mostly copied from the \RISCV{} backend of CompCert.

\paragraph{Outline}
We first review the work on which we directly build (\autoref{sec:background})
and present the key ideas of our work (\autoref{sec:key-ideas}).
%
%
Then we explain 
how we extended CompCert and
its correctness proof 
(\autoref{sec:extending-compcert}).
The following two sections detail the most interesting parts of our secure
compilation proof: back-translation (\autoref{sec:back-translation}) and
recomposition (\autoref{sec:recomposition}).
%
Then we present the assumption that the result of back-translation compiles and
how we thoroughly tested it (\autoref{sec:compiling-bt}).
We put these together into our secure compilation theorem
(\autoref{sec:top-level-theorem}) and then present our lower-level, unverified
capability backend (\autoref{sec:caps}).
We discuss related work (\autoref{sec:related})
before concluding with future work (\autoref{sec:future}).
Finally, the appendices include details that we had to cut for space.

\section{Background}
\label{sec:background}

In this section we briefly review the \rscdcmd{} secure compilation criterion of
Abate~\ETAL~\cite{AbateABEFHLPST18} as well as their high-level proof structure
for this criterion, since we make use of both in this paper.

But first, we warm up by reviewing the compiler correctness properties this
proof structure makes use of.
For this we assume that both the source language (for our security proof this is Clight)
and the target language (\RISCV assembly) are given trace-producing semantics.
CompCert traces are composed of events recording the calls the whole program
makes to system calls performing IO and the results they return.
A special event $\undefi{(k)}$ terminates the trace if undefined behavior is encountered
by compartment $k$ (where $k$ is something we added, and which we omit where it is irrelevant).
We further extend these traces to cross-compartment calls and returns
(\autoref{fig:events} in \autoref{sec:extending-compcert}).
Because the criterion of Abate~\ETAL~\cite{AbateABEFHLPST18} focuses
on safety properties~\cite{LamportS84} we only consider finite prefixes of traces.
We write $\src{W_S} \semsrc m$ when the whole source program $\src{W_S}$ can
produce the finite trace prefix $m$; and analogously $\trg{W_T} \semtrg m$ when
the whole target program $\trg{W_T}$ can produce $m$.
%
The compiler correctness guarantee of CompCert states that if a compiled
whole program can produce a trace prefix $m$ (\IE $\cmp{\src{W_S}} \semtrg m$) then the original
source program can produce a related trace $m' \preceq m$
(\IE  $\src{W_S} \semsrc m'$), where the relation $m' \preceq m$ is defined as
$m' = m$ when $\undefi \not\in m'$, and as $m'_0 \cdot m_1 = m$ when
$m' = m'_0 \cdot \undefi{(k)}$ for some $k$.
Here~``$\cdot$'' denotes concatenation and
$m_1$ is a completely arbitrary trace suffix that the correctly compiled
program is allowed to produce when the source program encounters undefined behavior,
which can lead to security vulnerabilities.

\begin{figure*}
\centering
\begin{tikzpicture}[auto]
  \node(cPi) {$\trg{p} : (\trg{C_T} \linktrg \cmp{\src{P}}) \semtrg m$ };
  \node[right = of cPi] (CP1t) { $(\cmp{\src{C_S}} \linktrg~ \cmp{\src{P'}}) \semtrg m$ };
  \node[align = left, above = of cPi] (CP1s) {
    $\begin{array}{r}
    \back{(I,\trg{p}, K)} ~= \src{C_S} \linksrc \src{P'}\\
    (\src{C_S} \linksrc \src{P'}) \semsrc m
    \end{array}$
  };
  \node[right = of CP1t, xshift=1.5em] (CPt) { $(\cmp{\src{C_S}} \linktrg~ \cmp{\src{P}}) \semtrg m$ };
  \node[above = of cPi, xshift=12em, yshift=1.5em] (prec) { $m' \preceq_{K\setminus\ii{CK}} m$ };
  \node[above = of CPt, xshift=0em] (CPs) { $(\src{C_S} \linksrc \src{P}) \semsrc m' \wedge m' \preceq m$};

  \draw[->] (cPi.90) to node {\hyperref[def:back-translation]{\em 1 Back-translation}} (CP1s.-90);

  \draw[->] (CP1s.-45) to node [align=left,font=\itshape,xshift=-0.25em,yshift=-0.3em]{\hyperref[def:fcc]{2 FCC}} (CP1t.180);

  \draw[->] (CP1s.-8) to node [right,yshift=-0.5em, xshift=3em]{\em \hyperref[def:blame]{5 Blame}} (prec.180);
  \draw[->] (CPs.140) to node {} (prec.0);

  \draw[->] (CP1t.0) to node [below,xshift=0em,yshift=-0.5em]{\em \hyperref[def:recomposition]{3 Recomposition}} (CPt.180);
  \coordinate [below=1.5em of CP1t,xshift=0em] (compoint1);
  \coordinate [below=1.5em of CPt,xshift=-4em] (compoint2);
  \draw[-] (cPi) to (compoint1);
  \draw[-] (compoint1) to (compoint2);
  \draw[->] (compoint2) to (CPt.220);

  \draw[->] (CPt.90) to node [right,align=left,font=\itshape]{\hyperref[def:bcc]{4 BCC}} (CPs.-90);

  \node[left = of CP1s, xshift=1.7em] () {
    $\begin{array}{c}
       \text{\color{gray} Source}\\
       \text{(Clight)}
     \end{array}$
   };
  \node[left = of cPi, xshift=1.5em,yshift=0.2em] () {
    $\begin{array}{c}
       \text{\color{gray} Target}\\
       \text{(\RISCV)}
     \end{array}$
   };
\end{tikzpicture}
\vspace{-1.2em}
\caption{The high-level proof structure for \rscdcmd
  of Abate~\ETAL\citeFull{AbateABEFHLPST18}{AbateABEFHLPSTT18ext}}
\label{fig:rsc-dc-md-proof}
\end{figure*}

Instead of compiling only whole programs though, we assume separate compilation---as proposed
by Kang~\ETAL~\cite{KangKHDV15} and implemented in CompCert since version
2.7---and separately compile a source program $\src{P}$ and a context $\src{C}$,
which are intuitively both formed of linked compartments, and
which can be linked together to produce a whole program both before
compilation using source linking ($\linksrc$), and after compilation using
target linking ($\linktrg$).
%
Using these concepts we can define the correctness of a compiler $\downarrow$
like CompCert or our variant SECOMP as follows:

\begin{definition}[Backward Compiler Correctness (BCC)]\label{def:bcc}
\[
  \forall \src{C}~\src{P}~m.~(\cmp{\src{C}} \linktrg~ \cmp{\src{P}}) \semtrg m \Rightarrow
    \exists m'.~(\src{C} \linksrc \src{P}) \semsrc m' \wedge m' \preceq m
\]
\end{definition}

In CompCert this backward compiler correctness (BCC) definition is proved by
forward simulation~\citeFull{Leroy09b}{BeringerSDA14},
so one also obtains a forward compiler correctness (FCC)
result, that the \rscdcmd proof structure of Abate~\ETAL~\cite{AbateABEFHLPST18}
also makes use of. Here, instead of obtaining a related trace prefix they instead
assume that the prefix one starts from in the source does not end with undefined behavior:

\begin{definition}[Forward Compiler Correctness (FCC) \cite{AbateABEFHLPST18}]\label{def:fcc}
\[
  \forall \src{C}~\src{P}.~ \forall m \not\ni \undefi.~
  (\src{C} \linksrc \src{P}) \semsrc m \Rightarrow (\cmp{\src{C}} \linktrg~ \cmp{\src{P}}) \semtrg m
\]
\end{definition}

All variants of C compiler correctness, including the two above, completely give
up on the whole program after it encounters undefined behavior.
To mitigate this issue, Abate~\ETAL~\cite{AbateABEFHLPST18} propose a secure
compilation notion that restricts the scope of undefined behavior to the
compartments that encounter it. 
Such compromised compartments can only influence other compartments via controlled
interactions respecting their interfaces and the other abstractions of the
source language (\EG the stack discipline on calls and returns).
Moreover, to model dynamic compromise the scope of undefined behavior is also
restricted temporally, by still providing protection to each compartment up to
the point in time when it encounters undefined behavior.

Abate~\ETAL~\cite{AbateABEFHLPST18} formalize this intuition as an iterative game
in which at each step some (initially empty)
set of compartments $\ii{CK}$  is already compromised and tries
to attack the remaining uncompromised compartments $K{\setminus}\ii{CK}$, for some
set of compartment identifiers $K$ defined in the original compartmentalized
program with global interface $I$, capturing all procedure imports and exports.
In each step, the uncompromised compartments are linked together into a source
program $\src{P}$ with interface $\lfloor I \rfloor_{K\setminus\ii{CK}}$, and then $\src{P}$
is compiled and linked with a target context $\trg{C_T}$, which puts together
the compromised compartments and which has interface $\lfloor I \rfloor_{\ii{CK}}$.
The guarantee obtained at each step in this 
game is formalized as a property they call \rscdcmd, which is defined then explained
below:\footnote{First time readers can also skim the remaining technical
  definitions in this section and focus on the intuitive explanations and the
  graphical representation from \autoref{fig:rsc-dc-md-proof}.  The paper of
  Abate~\ETAL~\cite{AbateABEFHLPST18} provides a gentler introduction to this
  proof structure. }

\begin{definition}\label{def:rsc-dc-md}
A compilation chain satisfies {\em Robustly Safe Compilation with
  Dynamic Compromise and Mutual Distrust (\rscdcmd)} if
there exists a back-translation function $\uparrow$
that takes 
interface $I$,
a target execution $\trg{p}$ producing a 
trace prefix $m$,
and a compartment identifier $k$, and generates a source compartment such that
\[
\begin{array}{l}
\forall K {\subseteq} \ii{CompIds}.~
\forall I {:} \ii{Interface}(K).~
\forall \ii{CK} {\subseteq} K.~
\forall \trg{C_T} {:} \lfloor I \rfloor_{\ii{CK}}.~
\forall \src{P} {:} \lfloor I \rfloor_{K\setminus\ii{CK}}.~\\
\forall m \not\ni \undefi.
  \forall \trg{p} : (\trg{C_T} {\linktrg} \cmp{\src{P}}) \semtrg m.~\\
\exists \src{C_S} {:} \lfloor I \rfloor_{\ii{CK}}.~
  \src{C_S} {=} \underset{k{\in}\ii{CK}}{\biglinksrc}\back{(I,\trg{p},k)} \land~
  \exists m'. (\src{C_S} {\linksrc} \src{P}) {\semsrc} m' \land
  m' {\preceq_{K{\setminus}\ii{CK}}} m
\end{array}
\]
\end{definition}

The premise on the first two lines 
states that the compound program
$\trg{C_T} \linktrg \cmp{\src{P}}$ has an execution $\trg{p}$ in the target
language producing a trace prefix $m$, which does not end with an undefined
behavior event
(\IE for a trace $m \cdot \undefi(k')$ one looks only at the prefix $m$).
The conclusion makes a step towards providing an explanation for $m$ with
respect to the source language semantics.
For this it calls the back-translation function $\uparrow$ on each of the
compromised compartments $k{\in}\ii{CK}$ and it links together the generated source
compartments to obtain a source context $\src{C_S}$ with interface
$\lfloor I \rfloor_{\ii{CK}}$.
%
The \rscdcmd property says that the
obtained context $\src{C_S}$ linked with the original source program
$\src{P}$ can produce a trace $m'$ that is related to $m$ by the formula
$m' \preceq_{K\setminus\ii{CK}} m$.
This is a variant of the $\preceq$ relation from BCC that ensures that
$m' = m$ when $\undefi \not\in m'$, and that $m'_0 \cdot m_1 = m$ when
$m' = m'_0 \cdot \undefi{(k)}$ for some {\em uncompromised} compartment
$k {\in} K{\setminus}\ii{CK}$.
Intuitively either the whole target 
prefix $m$ can be explained by an
execution in the source language, in which case we are done;
or the compromised compartments have
found a way to use the interface in the source language to trigger an undefined behavior
in one of the (so far) uncompromised compartments $k {\in} K{\setminus}\ii{CK}$.
In this second case, Abate~\ETAL~\cite{AbateABEFHLPST18} will apply $\rscdcmd$
again to an extended set of compromised compartments $\ii{CK} {\cup} \{ k \}$.
Because the semantics is determinate~\citeFull{Engelfriet85}{milner82,ChevalCD13}, with each
iterative application of \rscdcmd the execution is ``rewound'' along the
original trace prefix $m$ and longer and longer prefixes of $m$ are explained in
the source, until the whole $m$ is explained in terms of the source semantics
and a sequence of dynamic compartment compromises.

For proving secure compilation this iterative aspect is less interesting though,
and it basically suffices to show \rscdcmd~\cite{AbateABEFHLPST18}.
For this, Abate~\ETAL~\cite{AbateABEFHLPST18} propose the high-level proof
structure from \autoref{fig:rsc-dc-md-proof} that involves compiler
correctness (the FCC and BCC properties above) and three additional security-related
properties: back-translation, recomposition, and blame.
The high-level proof starts by back-translating a global interface $I$ and a
target execution $\trg{p}$ producing a trace prefix $m$ repeatedly to generate
each of the compartments $k \in K$ of a whole source program producing $m$.

\begin{definition}[Back-translation]\label{def:back-translation}
There exists a 
function $\uparrow$ s.t. 
\[
\begin{array}{l}
\forall K.~ 
\forall I.~ 
\forall \trg{W_T} {:} I.~
\forall \ii{CK} {\subseteq} K.~
\forall m \not\ni \undefi.
  \forall \trg{p} : \trg{W_T} \semtrg m.~\\
  \exists \src{C_S} {:} \lfloor I \rfloor_{\ii{CK}}.~
  \exists \src{P'} {:} \lfloor I \rfloor_{K\setminus\ii{CK}}.~
  \src{C_S} {\linksrc} \src{P'} {=} \underset{k\in K}{\biglinksrc}\back{(I,\trg{p},k)} \land~
  (\src{C_S} {\linksrc} \src{P'}) {\semsrc} m
\end{array}
\]
\end{definition}

Using the back-translation function $\uparrow$ to generate a {\em whole} source
program $\src{C_S} \linksrc \src{P'}$ not only allows the conclusion of
\autoref{def:back-translation} to be stated in terms of the usual operational
semantics of whole programs ($(\src{C_S} \linksrc \src{P'}) {\semsrc} m$), but 
also allows Abate~\ETAL~\cite{AbateABEFHLPST18} to compile this whole program and
make use of FCC in step 2 from \autoref{fig:rsc-dc-md-proof} to obtain that
$(\cmp{\src{C_S}} \linktrg~ \cmp{\src{P'}}) \semtrg m$.
Then, in step 3 they recompose the compartments from this execution with the ones
from original execution $(\trg{C_T} \linktrg \cmp{\src{P}}) \semtrg m$ to obtain the
execution $(\cmp{\src{C_S}} \linktrg~ \cmp{\src{P}}) \semtrg m$.

\begin{definition}[Recomposition]\label{def:recomposition}
\(\forall K.~ 
\forall I.~ 
\forall \ii{CK} {\subseteq} K.\)
\[
\begin{array}{l}
\forall \trg{C_T}, \trg{C_T'} : \lfloor I \rfloor_{\ii{CK}}.~
\forall \trg{P_T}, \trg{P_T'} : \lfloor I \rfloor_{K\setminus\ii{CK}}.~
\forall m. \\
(\trg{C_T} \linktrg \trg{P_T}) \semtrg m \land
(\trg{C_T'} \linktrg~ \trg{P_T'}) \semtrg m \Rightarrow
(\trg{C_T'} \linktrg~ \trg{P_T}) \semtrg m
\end{array}
\]
\end{definition}

Step 4 
uses BCC to turn target execution
$(\cmp{\src{C_S}} \linktrg~ \cmp{\src{P}}) \semtrg m$ back
into source execution $(\src{C_S} \linksrc \src{P}) \semsrc m'$, where the
relation $m' \preceq m$ accounts for the possibility of undefined behavior in
$\src{C_S} \linksrc \src{P}$.
The context $\src{C_S}$ is, however, generated by the back-translation and has
no undefined behavior along the trace $m$, which one shows in step 5 of the proof.
So if there is an undefined behavior in $m'$ then this can only be blamed on an
as-yet-uncompromised compartment $k \in K{\setminus}\ii{CK}$, as required by the
conclusion of \rscdcmd from \autoref{def:rsc-dc-md}.

\begin{definition}[Blame]\label{def:blame}
$\forall K.~ 
\forall I.~ 
\forall \ii{CK} {\subseteq} K.
\forall \src{C_S} {:} \lfloor I \rfloor_{\ii{CK}}.~
\forall \src{P}, \src{P'} {:} \lfloor I \rfloor_{K\setminus\ii{CK}}.$
\[
\forall m.(\src{C_S} \linksrc \src{P'}) \semsrc m \land
(\src{C_S} \linksrc \src{P}) \semsrc m' \wedge m' \preceq m \Rightarrow
m' \preceq_{K\setminus\ii{CK}} m.
\]
\end{definition}

\section{Key ideas}\label{sec:key-ideas}

\subsection{Compartment model}\label{sec:key-ideas:comp-interface}


Compartmentalization~\citeFull{GudkaWACDLMNR15, wedge_nsdi2008, Kilpatrick03}{VasilakisKRDDS18}
allows developers to structure large programs into mutually
distrustful compartments that have limited privileges, are isolated from each
other, and can only interact in a controlled way.
In this work, we adopt a model that statically partitions C programs into compartments.
Every C definition 
of a procedure or a global variable
belongs to a single compartment.
%

Any block of memory belongs to the compartment that allocated it and
compartments do not share memory: each block can only be accessed by the code of
the compartment it belongs to.
Instead, all interactions between compartments must happen via
\emph{cross-compartment calls and returns} that respect the
\emph{interfaces} provided by the programmers:
%
each compartment \(\mathtt{C}\) comes with a set of exported procedure declarations (\IE{}
which procedures it makes available to other compartments), written \(\mathtt{C.exports}\),
and a set of imported procedure declarations (\IE{} which procedures it uses from other compartments),
written \(\mathtt{C.imports}\).
Compartments must respect these compartment interfaces;
otherwise they trigger undefined behavior, whose scope is restricted by our
secure compilation property to just the offending compartment.
%
Moreover, in our model
compartments can only pass each other scalar values as procedure call
arguments and return values.
%
We introduced this last restriction for two reasons: first, a compartment cannot
use pointers from another compartment, which defeats the purpose of passing
pointers in most cases.
More importantly, passing pointers to other compartments would require recording
these pointers on the trace, which would significantly complicate
back-translation, recomposition, and if also done for pointers to dynamically
allocated memory also compiler correctness (see \autoref{sec:future}).


The compartments also interact with an external environment using \emph{system
  calls}.\footnote{Our ``system calls'' correspond in CompCert to
  the ``external'' functions that do not get resolved to actual C source code during linking.
  Such functions are implemented in lower-level, trusted libraries (like libc)
  and may include operating system calls.}
These are special, privileged procedures whose semantics is
axiomatized in CompCert
and that may generate some events.
Example of these system calls include volatile memory operations, calls to the
heap allocator, or input and output (e.g., reading from the console).
The system calls do not belong to a compartment; instead
calling them is considered a special kind of internal
call that can only change the calling compartment's memory.
Following the principle of least privilege, by default compartments do not have
access to system calls beyond safe operations like memory allocation and freeing.
%
Instead, the interfaces specify, for each compartment, which system calls they are allowed
to use.

\subsection{Adding compartments to CompCert}\label{sec:key-ideas:secure-sem}%
\label{sec:key-ideas:riscv}

\paragraph{Extension to CompCert's languages}
Following the ideas above, we extend all of CompCert's 10 languages, from
C to \RISCV{} assembly, adding syntax describing the compartment breakup and interfaces
and semantic checks to ensure all compartments respect these interfaces.
As explained above, failing a check triggers undefined behavior for the
offending compartment.
%
%
In particular, we update
memory operations to take an additional compartment argument,
ensuring compartments cannot access other compartments' memory.
Also, at every point where control could pass to another compartment
(calls and returns at higher levels, jumps at the lowest level)
we add a check that the control transfer respects the interfaces
and that compartments only pass each other scalar values.
Compilation preserves the program's interfaces and linking two partial programs
requires that they have compatible interfaces.

To prove secure compilation (\autoref{def:rsc-dc-md} from
\autoref{sec:background}), we also extended the CompCert
trace model with two new events:
\(\mathtt{Event\_call}\) and \(\mathtt{Event\_return}\).
These events are generated by cross-compartment calls and returns (or the
equivalent jumps in \RISCV{} assembly), and 
record enough
information on the traces to be able to prove recomposition and back-translation.
%
Since call and return events must not be disturbed by optimization, 
we disallow cross-compartment tail-call optimization
and inlining, as those would substantially change the way compartments interact
(\EG would require merging stack frames belonging to different
compartments).

%
%
At the \RISCV{} level, implementing secure compartments required even more care.
Without proper protection, adversarial code could make use of the unstructured
control-flow inherent to the assembly language to break the compartment abstraction.
For instance, an attacker could try to jump 
to code to which it shouldn't have access.
%
We modified CompCert's \RISCV{} assembly semantics to prevent this kind of attack, by
protecting the compartment abstraction and interfaces.
To do so, we observe that calls and returns are only implemented by the compiler
using specific instructions---jump-and-link for calls, and indirect jumps for
returns---so we forbid all other instructions from changing compartments.
Then, when an execution encounters such a jump and attempts to switch to
another compartment, we make use of the interfaces and of a newly added shadow
stack to decide whether the switch is allowed.
If the instruction is a jump-and-link, then the semantics checks whether it is
an allowed call according to the interfaces, and then records the return address
and the stack pointer on the shadow stack.
To decide which indirect jumps to allow to return to a different compartment,
we inspect the top of the shadow stack, and
make sure that the compartment performing the jump is returning to the right
address and has correctly restored the caller's stack pointer.
This prevents a malicious compartment from returning using the wrong
return address or confusing the caller about its stack.
More details about our changes to the semantics are discussed in \autoref{sec:extending-compcert}.


\paragraph{Correctness proofs}
We updated all of CompCert's 19 passes and the simulation proofs showing FCC
(\autoref{def:fcc}) to account for the addition of compartments and the new trace model,
which by determinacy implies BCC (\autoref{def:bcc}).
The updated proofs mainly rely on the compartment information being correctly
preserved by compilation, \EG{} procedures do not change compartment, memory
blocks that belong to different compartments are not merged, \ETC{}

Adapting CompCert's compiler correctness Coq proof to account for our
changes was a substantial amount of work.
We wanted to change the proof as little as possible, but since CompCert is a realistic
compiler, 
it was not always obvious from the start how best to do this.
Several times, we made design decisions that seemed adequate,
but that turned out to be inadequate much later
(\EG choosing at which precise step to insert a given check),
when we discovered that they interacted poorly with some particular
compilation pass (\EG intra-compartment inlining or tail-call optimization) or
language (\EG \RISCV assembly).
These issues often did not affect the correctness of the
compiler, but made the proofs much more difficult, so we had to backtrack and
find alternative ways to structure the changes so as to keep the proofs simple.

In the end we found elegant ways of adapting CompCert's compiler correctness Coq
proof to account for all the changes above and proved FCC and BCC.
Yet, while compiler correctness is an important part
of our security proof, it is definitely not sufficient by itself
(\autoref{sec:intro}). In the remainder of this section we discuss the
other components of the security proof (\autoref{sec:background}).
%

\subsection{Back-translation from RISC-V to Clight}\label{sec:key-ideas:back-translation}

In our setting, the first step of the proof structure from
\autoref{sec:background} is to back-translate a finite
\RISCV{} execution prefix into a \clight{} program that produces the same trace prefix.
We show constructively that given a (whole) compartmentalized \RISCV{} program
and a finite trace prefix of that program, there exists a (whole) compartmentalized
\clight{} program with the same interface that can also produce the same trace prefix.
%
%
Back-translation resembles compilation, but for
RSP~\citeFull{AbateBGHPT19, difftraces, PatrignaniG21}{PatrignaniG17} and variants
like the one we consider, the program obtained by back-translation only needs to
preserve \emph{one single finite trace prefix}, not every possible execution of
the original program.

Based on this observation, prior works~\cite{AbateABEFHLPST18,secureptrs,
  PatrignaniC15, BusiNBGDMP21, akram-capabileptrs, AgtenSJP12}
use a simple back-translation from a trace prefix to a program. Each procedure of the
program consists of a loop over a counter, 
which records how far the trace has been executed.
The body of the loop is a switch over the counter value;
the \(n\)th case of the switch contains code that will 
produce the \(n\)th event of the trace.
%
%
Proving such a back-translation correct can usually be done in two steps~\cite{AbateABEFHLPST18}: first,
one proves that all traces generated by a target program satisfy a \emph{well-formedness}
condition, and then, that every well-formed trace can be back-translated to a
program that produces that same trace.
We adapt this back-translation and proof technique to our setting, but to do so,
we need to make our events more informative and devise a novel
notion of well-formedness of traces made out of these informative events through an
intermediate language.

First, the events we use in SECOMP (introduced informally in
\autoref{sec:key-ideas:riscv} and detailed in \autoref{fig:events}
in \autoref{sec:extending-compcert})
do not contain enough information to
directly convert each trace into a \clight{} statement.
In particular, they do not capture all the information necessary to obtain a
back-translated program that produces the same events for system calls:
%
if a \RISCV{} system call produces an event with memory
\(\mathtt{M}\) and current compartment \(\mathtt{C}\), then for the back-translated
Clight program to produce the same event with memory \(\mathtt{M'}\) and current compartment \(\mathtt{C}\),
the CompCert-style axiomatization of system calls
requires us to prove that \(\mathtt{M}\) and \(\mathtt{M'}\)
are related by some memory injection~\cite{LeroyB08}
that is defined at least on the public symbols of \(\mathtt{C}\).
Put simply, this means that the semantics of system calls is allowed to depend
on the content of the calling compartment's global buffers.
Since we restrict our semantics to ensure global buffers only contain scalars when
calling system calls, this effectively means that \(\mathtt{M}\) and \(\mathtt{M'}\)
must have the same values in \(\mathtt{C}\)'s global buffers.\rb{Could
add a couple of words here to make more explicit the connection with the
back-translation of the RISC-V buffers after shortening below}\jt{I think this is fine}
Yet, the SECOMP events do not include the content of global buffers.

This motivates introducing more \emph{informative events}
satisfying two requirements: (1) a
\RISCV{} program always produces a well-formed trace of informative events, and (2)
each informative event directly translates into
a 
\clight{} statement.
%
Using these informative events, we define a novel notion of \emph{well-formed informative trace}
and a back-translation from such traces to \clight{} programs, and we
significantly extend the technique of Abate~\ETAL~\cite{AbateABEFHLPST18} to
prove the correctness of this back-translation.
%
%
Informative events 
augment each system call event (and also cross-compartment call or return events) with 
a list of the changes to the global buffers since the last informative event. 
Each of these changes is called a \emph{memory delta}, written \(\delta\), and a
list of these is written \(\Delta\), and is ordered chronologically.
Our back-translation 
uses these deltas to generate Clight code performing
the same changes to global buffers before performing the system call.

First, we define a novel notion of well-formedness of a trace of informative events.
To do so, we define a new intermediate language\footnote{We see our
  back-translation function as a compiler from traces to Cminor programs.}
with a semantics that characterizes the well-formedness of such traces.
Informally, informative traces can be seen as the programs of this language, and
the step relation executes these traces by emitting the first informative event
of the trace and updating the state according to the event.
More precisely, in this language, states \(s\) are triples that record the currently executing procedure, a memory,
and a cross-compartment call stack; the step relation \(s \xrightarrow{\alpha} s'\)
relates the states \(s\) and \(s'\) and produces an informative event \(\alpha\).
The rules of the step relation additionally record the conditions necessary for the
back-translated code to be proved correct.
In particular, if \(s \xrightarrow{\alpha} s'\), then applying the deltas \(\Delta\)
from \(\alpha\) to the memory \(\mathtt{M}\) of \(s\) produces the memory \(\mathtt{M'}\) of \(s'\)
(\IE \(\mathtt{mem\_delta\_apply}~\Delta~{\mathtt M} = {\mathtt M'}\)).


We say that a trace of informative events is \emph{well-formed} when it is
produced by the reflexive transitive closure of this step relation.
We prove that for any trace prefix \(m\) produced by a \RISCV{} program,
there exists a well-formed informative trace \(\overline{\alpha}\)
that projects to \(m\) by removing the additional data recorded in informative events.
%
This allows us to then define a back-translation function on informative traces,
and completely forget about the \RISCV{} semantics when proving the correctness of this back-translation.

This back-translation function operates in the standard way explained above,
except that it also uses the memory deltas to generate code that writes the
right values to the global buffers before performing a system call.
We prove the correctness of the back-translation by induction on the intermediate language
execution, using the information provided by the step relation.
In order to prove that system calls at the \clight{} level generate the same events,
we maintain an invariant on permissions of global buffers, and show
that it can be used together with memory deltas to build a memory injection
that allows us to use the axiomatization of system calls from CompCert.
We give more details of this 
in \autoref{sec:back-translation}.

\subsection{Recomposition for RISC-V compartments}\label{sec:key-ideas:recomposition}

Recomposition is essential for the proof structure of \autoref{sec:background}, as it allows to
replace the arbitrary \RISCV{} context with which we started, with a context that was
obtained by back-translation and then compiled.
More concretely, starting from two whole target programs \(\trg{W_{1}} =
\trg{C_{1}} \linktrg \trg{P_{1}} \) and \(\trg{W_{2}} = \trg{C_{2}} \linktrg
\trg{P_{2}}\) that can execute to produce the same trace \(m\) and that are each
split into a {\em program side} ($\trg{P_{1}}$ respectively $\trg{P_{2}}$) and a
{\em context side} ($\trg{C_{1}}$ respectively $\trg{C_{2}}$), recomposition
gives us that the whole program \(\trg{W_{3}} = \trg{C_{1}} \linktrg
\trg{P_{2}}\) produces the same trace \(m\).
%

The key intuitions behind this kind of proof~\citeFull{AbateABEFHLPST18, secureptrs,
  AbateBGHPT19, JuglaretHAEP16, PatrignaniC15, PatrignaniASJCP15,
  akram-capabileptrs, PatrignaniG21}{PatrignaniG17}
%
are as follows: (1) because of determinacy the
internal behavior of a compartment only depends on its internal state, and on
all the information it received from other compartments, or from system calls;
(2) our extended trace model is informative enough to capture all information that is
exchanged between compartments or obtained from system calls.
Because of this, we can relate the execution of \(\trg{W_{3}}\) to that of \(\trg{W_{1}}\) when
executing in a compartment of the context, and to that of \(\trg{W_{2}}\) otherwise:
the internal state of the compartments of the context part agree in the execution
of \(\trg{W_{3}}\) and that of \(\trg{W_{1}}\), and the same holds for \(\trg{W_{3}}\),
\(\trg{W_{2}}\), and the internal state of the compartments in the program part.
This is preserved by silent steps, because they only depend on that same internal state;
and because the trace events record every information exchanged, even when switching side,
non-silent-step also preserve this information.



Formalizing this idea in order to prove recomposition for our \RISCV{} semantics
extended with compartments is highly complex, as it relies on many
low-level details of the \RISCV{} semantics.
For this reason, we propose a generic proof technique that elegantly splits
recomposition into several self-contained parts.
Our technique introduces eight novel CompCert-like simulation diagrams (described in detail
in \autoref{sec:recomposition}) that provide a structured way
to think and reason about recomposition, explicitly separating the definition of
invariants, the reasoning about internal steps, and the reasoning about events
and cross-compartment communication.
Together, the diagrams imply the existence of a novel \emph{three-way recomposition simulation} that
itself implies the recomposition theorem (\autoref{def:recomposition}).

\begin{figure*}
  \centering

\renewcommand{\thesubfigure}{\arabic{subfigure}}

\begin{subfigure}{0.125\textwidth}
  \centering
  \begin{tikzpicture}
    \tikzstyle{every node}=[BlueViolet]
    \tikzstyle{every path}=[BlueViolet]

    \node[draw,shape=circle,fill=BlueViolet,inner sep=0.5pt](ini1) {};
    \node[left = 0.1pt of ini1, inner sep=0.5pt] (textini1) {\(s_{1}\) initial};

    \node[draw,shape=circle,fill=BlueViolet,inner sep=0.5pt,below = 0.75cm of ini1] (ini2) {};
    \node[left = 0.1pt of ini2, inner sep=0.5pt] (textini2) {\(s_{2}\) initial};

    \node[draw,shape=circle,fill=BlueViolet,inner sep=0.5pt,below = 0.75cm of ini2] (ini3) {};
    \node[left = 0.1pt of ini3, inner sep=0.5pt] (textini3) {\color{PineGreen}\(s_{3}\) initial};

    \draw[PineGreen] (-1.25,0.35) rectangle (0.10,-1.85);
    \node[right = 0.2cm of ini3, inner sep=0.5pt] (m) {\(\color{PineGreen}\mathcal{R}\)};
  \end{tikzpicture}
  \vspace{-0.5em}
  \caption{Initial states}\label{fig:threeway-simulationIni}
\end{subfigure}
\begin{subfigure}{0.225\textwidth}
  \centering
  \begin{tikzpicture}
    \tikzstyle{every node}=[BlueViolet]
    \tikzstyle{every path}=[BlueViolet]

    \node[draw,shape=circle,fill=BlueViolet,inner sep=0.5pt](ini1) {};
    \node[left = 0.1pt of ini1, inner sep=0.5pt] (textini1) {\(s_{1}\)};

    \node[draw,shape=circle,fill=BlueViolet,inner sep=0.5pt,right = 1cm of ini1] (after1) {};
    \node[right = 0.12pt of after1, inner sep=0.5pt] (textafter1) {\(s_{1}'\)};

    \node[draw,shape=circle,fill=BlueViolet,inner sep=0.5pt,below = 0.75cm of ini1] (ini2) {};
    \node[left = 0.1pt of ini2, inner sep=0.5pt] (textini2) {\(s_{2}\)};

    \node[draw,shape=circle,fill=BlueViolet,inner sep=0.5pt,right = 1cm of ini2] (after2) {};
    \node[right = 0.12pt of after2, inner sep=0.5pt] (textafter2) {\(s_{2}'\)};

    \node[draw,shape=circle,fill=BlueViolet,inner sep=0.5pt,below = 0.75cm of ini2] (ini3) {};
    \node[left = 0.1pt of ini3, inner sep=0.5pt] (textini3) {\(s_{3}\)};
    \node[draw,shape=circle,fill=PineGreen,PineGreen,inner sep=0.5pt,right = 1cm of ini3] (after3) {};
    \node[right = 0.12pt of after3, inner sep=0.5pt] (textafter3) {\(\color{PineGreen}s_{3}'\)};

    \draw[->,shorten >= 2pt, shorten <= 2pt,thick] (ini1) to node [above] {\(a\)} (after1.-180);
    \draw[->,shorten >= 2pt, shorten <= 2pt,thick] (ini2) to node [above] {\(a\)} (after2.-180);
    \draw[->,shorten >= 2pt, shorten <= 2pt,color=PineGreen] (ini3) to node [above] {\(\color{PineGreen}a\)} (after3.-180);
    \node[above left = 0.12pt of after3, inner sep=0.5pt] (star3) {\(\color{PineGreen}*\)};

    \draw[thick] (-0.40,0.35) rectangle (0.30,-1.85);
    \node[left = 0.5cm of ini3, inner sep=0.5pt] (m) {\(\mathcal{R}\)};

    \draw[PineGreen] (0.85,0.35) rectangle (1.5,-1.85);
    \node[right = 0.52cm of after3, inner sep=0.5pt] (mm) {\(\color{PineGreen}\mathcal{R}\)};
  \end{tikzpicture}
  \vspace{-0.5em}
  \caption{Non-silent step}\label{fig:threeway-simulationA}
\end{subfigure}
\begin{subfigure}{0.275\textwidth}
  \centering
  \begin{tikzpicture}
    \tikzstyle{every node}=[BlueViolet]
    \tikzstyle{every path}=[BlueViolet]

    \node[draw,shape=circle,fill=BlueViolet,inner sep=0.5pt](ini1) {};
    \node[left = 0.1pt of ini1, inner sep=0.5pt] (textini1) {\(s_{1}\)};

    \node[draw,shape=circle,fill=BlueViolet,inner sep=0.5pt,right = 1cm of ini1] (after1) {};
    \node[right = 0.12pt of after1, inner sep=0.5pt] (textafter1) {\(s_{1}'\)};

    \node[draw,shape=circle,fill=BlueViolet,inner sep=0.5pt,below = 0.75cm of ini1] (ini2) {};
    \node[left = 0.1pt of ini2, inner sep=0.5pt] (textini2) {\(s_{2}\)};

    \node[draw,shape=circle,fill=BlueViolet,inner sep=0.5pt,right = 1cm of ini2] (after2) {};
    \node[right = 0.12pt of after2, inner sep=0.5pt] (textafter2) {\(s_{2}\)};

    \node[draw,shape=circle,fill=BlueViolet,inner sep=0.5pt,below = 0.75cm of ini2] (ini3) {};
    \node[left = 0.1pt of ini3, inner sep=0.5pt] (textini3) {\(s_{3}\)};
    \node[draw,shape=circle,fill=PineGreen,PineGreen,inner sep=0.5pt,right = 1cm of ini3] (after3) {};
    \node[right = 0.12pt of after3, inner sep=0.5pt] (textafter3) {\(\color{PineGreen}s_{3}'\)};

    \draw[->,shorten >= 2pt, shorten <= 2pt,thick] (ini1) to node [above] {\(\varepsilon\)} (after1.-180);
    \draw[double,shorten >= 2pt, shorten <= 2pt] (ini2) to node [above] {\(=\)} (after2.-180);
    \draw[->,shorten >= 2pt, shorten <= 2pt,color=PineGreen] (ini3) to node [above] {\(\color{PineGreen}\varepsilon\)} (after3.-180);
    \node[above left = 0.12pt of after3, inner sep=0.5pt] (star3) {\(\color{PineGreen}*\)};

    \draw[thick] (-0.40,0.35) rectangle (0.30,-1.85);
    \node[left = 0.5cm of ini3, inner sep=0.5pt] (m) {\(\mathcal{R}\)};

    \draw[PineGreen] (0.85,0.35) rectangle (1.5,-1.85);
    \node[right = 0.52cm of after3, inner sep=0.5pt] (mm) {\(\color{PineGreen}\mathcal{R}\)};
  \end{tikzpicture}
  \vspace{-0.5em}
  \caption{Silent step in first execution}\label{fig:threeway-simulationB}
\end{subfigure}
\begin{subfigure}{0.275\textwidth}
  \centering
  \begin{tikzpicture}
    \tikzstyle{every node}=[BlueViolet]
    \tikzstyle{every path}=[BlueViolet]

    \node[draw,shape=circle,fill=BlueViolet,inner sep=0.5pt](ini1) {};
    \node[left = 0.1pt of ini1, inner sep=0.5pt] (textini1) {\(s_{1}\)};

    \node[draw,shape=circle,fill=BlueViolet,inner sep=0.5pt,right = 1cm of ini1] (after1) {};
    \node[right = 0.12pt of after1, inner sep=0.5pt] (textafter1) {\(s_{1}\)};

    \node[draw,shape=circle,fill=BlueViolet,inner sep=0.5pt,below = 0.75cm of ini1] (ini2) {};
    \node[left = 0.1pt of ini2, inner sep=0.5pt] (textini2) {\(s_{2}\)};

    \node[draw,shape=circle,fill=BlueViolet,inner sep=0.5pt,right = 1cm of ini2] (after2) {};
    \node[right = 0.12pt of after2, inner sep=0.5pt] (textafter2) {\(s_{2}'\)};

    \node[draw,shape=circle,fill=BlueViolet,inner sep=0.5pt,below = 0.75cm of ini2] (ini3) {};
    \node[left = 0.1pt of ini3, inner sep=0.5pt] (textini3) {\(s_{3}\)};
    \node[draw,shape=circle,fill=PineGreen,PineGreen,inner sep=0.5pt,right = 1cm of ini3] (after3) {};
    \node[right = 0.12pt of after3, inner sep=0.5pt] (textafter3) {\(\color{PineGreen}s_{3}'\)};

    \draw[double,shorten >= 2pt, shorten <= 2pt] (ini1) to node [above] {\(=\)} (after1.-180);
    \draw[->,shorten >= 2pt, shorten <= 2pt,thick] (ini2) to node [above] {\(\varepsilon\)} (after2.-180);
    \draw[->,shorten >= 2pt, shorten <= 2pt,color=PineGreen] (ini3) to node [above] {\(\color{PineGreen}\varepsilon\)} (after3.-180);
    \node[above left = 0.12pt of after3, inner sep=0.5pt] (star3) {\(\color{PineGreen}*\)};

    \draw[thick] (-0.40,0.35) rectangle (0.30,-1.85);
    \node[left = 0.5cm of ini3, inner sep=0.5pt] (m) {\(\mathcal{R}\)};

    \draw[PineGreen] (0.85,0.35) rectangle (1.5,-1.85);
    \node[right = 0.52cm of after3, inner sep=0.5pt] (mm) {\(\color{PineGreen}\mathcal{R}\)};
  \end{tikzpicture}
  \vspace{-0.5em}
  \caption{Silent step in second execution}\label{fig:threeway-simulationC}
\end{subfigure}
  \vspace{-1em}
  \caption{Three-way simulation properties represented graphically}\label{fig:threeway-simulation}
  \vspace{-1em}
\end{figure*}
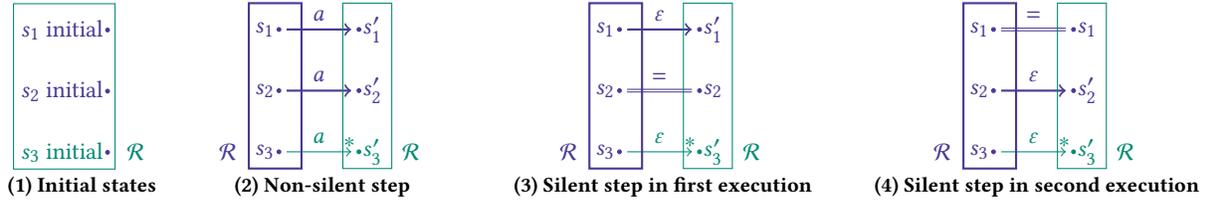

We define three-way recomposition simulation in the generic setting on labeled
transition systems \((S, \rightarrow)\) with \emph{initial} states:
\begin{definition}[Three-way recomposition simulation]
Given three labeled transition systems \(L_{1} = (S_{1}, \rightarrow_{1})\),
\(L_{2} = (S_{2}, \rightarrow_{2})\) and \(L_{3} = (S_{3}, \rightarrow_{3})\),
we say there exists a three-way recomposition simulation between \(L_{1}\),
\(L_{2}\), and \(L_{3}\) when there exists a relation \(\mathcal{R}\) between
states of \(L_{1}\), \(L_{2}\), and \(L_{3}\) that satisfies the properties
depicted in \autoref{fig:threeway-simulation}.
\end{definition}

For each property from \autoref{fig:threeway-simulation}, each row represents one of the 3 executions.
Arrows represent execution steps, and are annotated with either an event \(a\) or silence \(\varepsilon\).
We denote the reflexive transitive closure of the step relation with \(\to^{*}\).
We use thick, dark purple for the assumptions, and dark green for the conclusions we have to prove (including the existence
of states).
We denote equality via a double line.
We represent the simulation relation as a rectangle around the states it relates.
Property (1) states that the simulation relation \(\mathcal{R}\) is compatible with initial states.
Property (2) states that whenever the first two executions take a step from
related states producing the same observable event $a$, then so can the third one.
Properties (3) and (4) state that silent steps in either of the first two executions
preserve the simulation relation (and the third execution is allowed to take
  some silent steps too).

We prove that, given such a three-way simulation between the semantics of
programs \(W_{1}\), \(W_{2}\), and \(W_{3}\), then
\(W_{1} \leadsto m \wedge W_{2} \leadsto m \implies W_{3} \leadsto m\) (which
directly implies recomposition).
For this it is enough to follow both executions in \(W_{1}\) and \(W_{2}\),
and to apply the appropriate property, (2), (3), or (4) until all of \(m\) is
produced.\footnote{CompCert experts may note that we do not use a notion of
  decreasing measure or a notion of final states, since we are only concerned
  with finite execution prefixes.}


To define simpler to prove simulation diagrams in \autoref{sec:recomposition} we will
instantiate the relation \(\mathcal{R}\) above with the conjunction of three
relations, following ideas of El-Korashy\ETAL~\cite{akram-capabileptrs,
  secureptrs}: a \emph{strong} relation \(\sim\),
a \emph{weak} relation \(\equiv\), and a \emph{mixed} relation \(\mathcal{M}\).
%
Intuitively, the strong relation $\sim$ relates the internal state of the side being executed, the weak
relation $\equiv$ relates the internal state of the side not being executed, and the mixed relation \(\mathcal{M}\) relates the parts
shared between compartments (such as the stack).
Given three states \(s_{1}\), \(s_{2}\), \(s_{3}\) that are executing
a compartment that's taken from
\(W_{1}\), 
the strong relation \(\equiv\) relates
\(s_{1}\) 
to \(s_{3}\), and the weak relation \(\sim\) relates
\(s_{2}\) 
to \(s_{3}\).
Symmetrically, when the current compartment is taken from \(W_{2}\) then
\(s_{1} \sim s_{3}\) and \(s_{2} \equiv s_{3}\).
When switching to a compartment taken from the other side the relations are switched as well.

Compared to prior work~\cite{akram-capabileptrs,
  secureptrs}, a significant challenge in our proofs is that in our
realistic setting the three relations above are parameterized by two CompCert
\emph{memory injections}~\cite{LeroyB08}, one for each of the original executions.
Their role is to relate the memory of their respective execution to the memory in the
recomposed execution.
In particular, these injections do not relate the memory locations of the side that
doesn't correspond to their execution, and they are preserved by the simulation steps.
Essentially, both weak and strong relations are instantiated so that they relate
the memories of the runs according to these injections, and additionally the
strong relation also relates the content of the registers.
The mixed relation relates the content of the stack and cross-compartment stack,
and uses both memory injections.

\subsection{Blame for \clight semantics}\label{sec:key-ideas:security-proof:blame}


The blame theorem is the final proof step of \autoref{sec:background}
and shows that the back-translated
program $\src{C_S} \linksrc \src{P'}$ is free of
undefined behavior along the given trace prefix.
From this, it follows that any undefined behavior in $\src{C_S} \linksrc \src{P}$
must come from the original (partial) program $\src{P}$.
Blame relates the executions of two whole \clight programs
that produce the same trace prefix and
share a common set of compartments from $\src{C_S}$---their \emph{context
side}---linked with a pair of compatible (same public symbols, imports
and exports, etc.), but otherwise arbitrary
%
%
\emph{program sides} $\src{P}$ and $\src{P'}$, which supply the remaining compartments.
%
%
%
Intuitively, because the two executions produce
the same trace prefix, the shared context side
affects the two program sides
of the executions 
in equivalent ways, and any differences, including undefined behavior, must
originate in the different program sides.
%
%


Some of the intuitions of our blame proof are similar to the recomposition proof,
but carried over to \clight and to
a different type of simulation involving three partial program parts arranged
in two whole programs.
The key challenge of the blame proof 
lies in the definition of the
simulation invariants that relate the two executions. The shared trace prefix
forces both executions to run in sync: at any point in time, either the
shared context side is driving the two runs \emph{in lockstep}, or each
program side is running \emph{independently} from the other until an
observable event forces them to re-synchronize.
We can apply the ideas outlined above to prove a small number of
elementary simulation
results for an appropriate relation
$\mathcal{R}$, and use these to assemble a full proof of blame:

\begin{enumerate}[leftmargin=15pt,nosep]
\item $\mathcal{R}$ is preserved when \emph{both} whole programs take a
\emph{single step}, with both producing the \emph{same event}.
(Proved separately starting from the program side and the context side.)
\item $\mathcal{R}$ is preserved when \emph{one} of the whole programs
takes a 
\emph{silent step} from the \emph{program side} while the other stays put.
\end{enumerate}

On top of these stepwise results, we can build three preservation
properties on longer \emph{synchronized executions} of the two programs.

\begin{enumerate}[leftmargin=15pt,nosep]
\item $\mathcal{R}$ is preserved when \emph{both} whole programs
take a sequence of \emph{silent steps},
followed by a pair of \emph{synchronous steps}
producing the \emph{same event} on both sides. (Proved separately starting from the
program side and the context side.)
\item $\mathcal{R}$ is preserved after any \emph{shared trace prefix}
produced by a pair of executions of the two whole programs.
\end{enumerate}

Finally, synchronized executions allow us to reason about full program
runs, which result from the whole program taking steps until it is
unable to do so any more; we will simply refer to this as
a \emph{whole run}. After a whole program finishes execution, CompCert
can reason about its normal or abnormal termination by inspecting the
last state of the run. In particular, the semantics defines which
states are considered proper \emph{final states}; and all others are
considered \emph{stuck states}. A final state corresponds to
successful termination, and stuckness corresponds to undefined
behavior.
The following two key lemmas look at whole runs, the traces produced
by those runs, and the side that was in control at the end of the
execution (program or context) to blame the program side for any
differences.



\begin{enumerate}[leftmargin=15pt,nosep]

\item If $\mathcal{R}$ holds, one whole program runs with a trace $m$ and
the context side in control at the end, and the other whole program
runs with an extension of $m$ and terminates in a final state, then
the first whole program also ended in a final state.
%
\item If $\mathcal{R}$ holds, one whole program runs with a trace $m$, and
the other whole program runs with a \emph{strict extension} of the
trace, i.e., $m$ followed by a non-empty trace $m'$, then
the \emph{program side} of the first program was in control
at the end.
%
\end{enumerate}


The high-level blame proof 
follows in a relatively straightforward manner from the simulation lemmas for whole runs.
%
%
%
%
%
The main sources of complexity of the proof pertain to the
preservation of the blame invariants in the stepwise simulation lemmas
above. Notably, blame requires us to build and maintain asymmetric
memory injections that relate the contents of the context sides of a
pair of memories and their symbols. However, it cannot completely
disregard the program sides of the memories, which have to be
compatible, i.e., define the same public symbols, even if they are
different. For this reason, the injections must map all public symbols
in the whole programs, not just those in the shared context side.




\section{Extending CompCert with Compartments}
\label{sec:extending-compcert}


In this section, we detail how we added compartments to CompCert's languages,
including \RISCV{} assembly.
%

\paragraph{Memory model}
We reuse the block-based memory model of CompCert and extend it with compartments.
Each memory block belongs to a single compartment that is assigned at 
allocation and cannot be changed during the execution.
Memory operations (reads, writes, frees) are parameterized by the compartment
performing the operation, and fail when this compartment does not own the targeted
block.{}
This means that compartments cannot share memory or pass data other than by
performing calls and returns.

\paragraph{Calls and returns}
We extend CompCert's trace model to include two more events capturing compartment transitions 
(see \autoref{fig:events}). We write \(\overline{x}\) to denote a list of \(x\).
\begin{wrapfigure}[9]{L}[0pt]{0.25\textwidth}
  \(\arraycolsep=1pt
    \begin{array}{lcl}
      a &\mathrel{:=}&~\mathtt{Event\_syscall~name~\overline{v}~\colorbox{blue!30}{\(\overline{\mathtt{bs}}\)}~v~\colorbox{blue!30}{\(\overline{\mathtt{bs}}\)}} \\
                    &|& ~\mathtt{Event\_vload~m~ch~id~ofs~v} \\
                    &|&~\mathtt{Event\_vstore~m~ch~id~ofs~v} \\
                    &|&~\mathtt{Event\_annot~string~\overline{v}}\\
                    &|&\colorbox{blue!30}{\(\mathtt{Event\_call~cp~cp.id~\overline{v}}\)}\\
                    &|&\colorbox{blue!30}{\(\mathtt{Event\_return~cp~cp~v}\)}
    \end{array}
    \)
  \vspace{-1.0em}
  \caption{SECOMP events}\label{fig:events}
\end{wrapfigure}
The new events are highlighted: \(\mathtt{Event\_call~C~C'.f~args}\) captures a
cross-compartment call passing \(\mathtt{args}\) from compartment \(\mathtt{C}\)
to the procedure \(\mathtt{f}\) of compartment \(\mathtt{C'}\).
Similarly, \(\mathtt{Event\_return~C'~C~v}\) represents returning value
\(\mathtt{v}\) from compartment \(\mathtt{C'}\) to compartment \(\mathtt{C}\).

For all languages but \RISCV{}, we make use of the existing structure of
the semantics to implement these new events, so the only required change is to
add the events to the appropriate call and return transitions.
We also insert in the semantics dynamic checks to ensure calls
and returns conform to the interfaces and do not pass pointers.
%
%
%
The dynamic checks act as follows:
Internal calls (\EG a call from \(\mathtt{C.f}\) to \(\mathtt{C.g}\)) are always allowed.
System calls are allowed only if the interface allows it.
A cross-compartment call from \(\mathtt{C.f}\) to \(\mathtt{C'.g}\)
is allowed if (1) \(\mathtt{g} \in \mathtt{C'.exports}\);
(2) \(\mathtt{C'.g} \in \mathtt{C.imports}\); and
(3) all of the arguments are scalar values.
The dynamic checks for cross-compartment returns are similar: we
check that the returned value is a scalar. 
In all languages before \RISCV control-flow well-bracketedness
is ensured by the semantics.
%
\ifanon\else\ifcamera\else
We describe in more details on an example how calls and returns are executed in
SECOMP in \autoref{app:call-return}
\fi\fi

Lastly, we made all registers be caller saved, since on a cross-compartment call we
cannot trust the callee compartment to save and restore the caller's 
registers.\footnote{For simplicity our current implementation does this for all
  calls, not just cross-compartment ones, but it would be good to improve this
  aspect in the future.}
We also made the semantics invalidate non-argument registers on cross-compartment
calls and non-return registers on cross-compartment returns (by making them
undefined values), since recomposition requires all information
passed between compartments to be captured by the trace.

\paragraph{Changes to \RISCV}
As explained in \autoref{sec:key-ideas:riscv}, adding compartments to the
\RISCV{} assembly semantics required more extensive changes.
For a start we added a boolean flag to the jump-and-link instructions
$\mathtt{Pjal_{s}}$ and $\mathtt{Pjal_{r}}$, and to the indirect jump
instruction $\mathtt{Pj_{r}}$.
When this tag is \(\mathtt{true}\), the instruction can be used to attempt
cross-compartment calls using jump-and-link or returns using indirect jump,
but otherwise instructions are not allowed to change the current compartment.
Additionally, a change we did to the RISC-V semantics for recomposition
to hold is to make stack frames used to pass spilled arguments in cross-compartment
calls read-only.
This prevents a malicious compartment to exploit callbacks to modify the spilled
arguments it previously passed, for instance by reusing a pointer to a former
stack frame that contains such arguments.

Most importantly, our \RISCV{} semantics makes use of a \emph{cross-compartment
  shadow stack}, which is a list of records,
each containing a return address, a stack pointer, and a procedure signature.
Whenever a cross-compartment call occurs, the previous return address and stack
pointer, and the callee's signature are pushed on top of the shadow stack.
Whenever a cross-compartment return occurs, the semantics checks that the return targets the
right address and that the stack pointer has been correctly restored, before
executing the return.
The procedure signature on the shadow stack includes what register stores the
return value, which is used to check that no pointers are passed.
This shadow stack is used for abstractly specifying the
well-bracketedness~\cite{AndersonBLPT23} of cross-compartment control-flow, and
leaves lower-level backends the freedom to enforce this in various ways---for
instance, in \autoref{sec:caps} we describe an enforcement mechanism based on capabilities.
Finally, this shadow stack is not used for calls and returns inside a
compartment, so backends don't have to do anything special for these.

\paragraph{Buffer-based IO}
We also extended the IO model of CompCert from single-character-based to buffer-based IO.
Before our change CompCert modeled only very simple IO procedures.
System calls and their arguments and
return value, which must be scalars or pointers to globals, are recorded as events in the program trace.
The behavior of all system calls is described by a single high-level axiomatization
that enforces various generic properties, which are sufficient to support the compiler
correctness proof. In particular, system calls are required
to be \emph{determinate}~\citeFull{Engelfriet85}{milner82,ChevalCD13}, in the sense that 
any two calls with the same arguments and results have the same effect on memory,
and \emph{receptive}, meaning that any call might return an arbitrary result value
of the correct type.  While this is adequate to model single-character-based IO
procedures like {\tt getchar} and {\tt putchar}, it does not account properly for
calls that read or write memory as a side-effect, which are very common in real
C code.  For example, the system call {\tt read} takes as arguments the number of
bytes to read and a buffer address, stores bytes into the buffer,
and returns the actual count of bytes that were stored.
Two {\tt read} calls might return the same count but store different
values in memory (violating determinacy); moreover, the count cannot exceed
the requested number of bytes (violating receptivity). 

We address these limitations by extending CompCert's system call events to 
record any bytes loaded from or stored to global memory buffers (the highlighted
arguments to {\tt Event\_syscall} in \autoref{fig:events}).
We weaken determinacy to allow calls to store
different byte values even if they return the same result value,
and we weaken receptivity to put procedure-specific constraints on result values
and stored bytes. For the latter, we just require that the result and stored bytes
might be produced by a call to the procedure with \emph{some} environment and initial memory.
To validate our approach, we give detailed models for {\tt read} and {\tt write}
system calls, and show that they indeed satisfy the (weakened) properties.

\section{Back-Translation Proof Details}
\label{sec:back-translation}

\newcommand{\btgen}{BT\xspace}
\newcommand{\bevent}{\alpha\xspace}
\newcommand{\btrace}{\overline{\bevent}\xspace}
\newcommand{\mdelta}{\delta\xspace}

\newcommand{\bstate}{\mathtt{state}}
\newcommand{\bstep}{\mathtt{step}}
\newcommand{\bgenv}{\mathtt{ge}}
\newcommand{\mblock}{\mathtt{block}}
\newcommand{\mmemory}{\mathtt{mem}}
\newcommand{\mlist}{\mathtt{list}}
\newcommand{\mprop}{\mathtt{Prop}}

We now detail the challenges involved in adapting the back-translation proof of
Abate~\ETAL{}~\cite{AbateABEFHLPST18} to our setting.
As explained in \autoref{sec:key-ideas:back-translation}, the main difficulty
stems from the fact that when a system call in \RISCV{} generates an event, 
we have to prove that it is possible to generate the exact same event in \clight{}.
However, the axiomatization of CompCert's system calls does not allow us to do
this easily. Among the axioms CompCert gives us, determinacy, which states that executing
the system call in the same memory state yields the same result, is not sufficient.
There is indeed little hope of perfectly reproducing the \RISCV{} memory in \clight{}.

Instead, CompCert provides another useful axiom:
if a system call is executed in some memory \(\mathtt{M_{1}}\) resulting
in memory \(\mathtt{M_{1}'}\) while generating a trace, then if executed in some other memory
\(\mathtt{M_{2}}\) that \(\mathtt{M_{1}}\) injects into, the same system call with the same arguments results in
a memory \(\mathtt{M_{2}'}\) that \(\mathtt{M_{1}}'\) injects into, and crucially generates the same trace.
Yet using this axiom imposes another condition on the memory injection:
it must be defined at least on the public symbols of the environment (of the calling compartment),
which include global buffers.
%
%
This motivates our usage of informative events to record the content of these global
buffers inside memory deltas.

\paragraph{Informative events}
We consider 3 kinds of informative events:
\begin{enumerate}[leftmargin=15pt,nosep]
  \item $\mathtt{Icall~f~t~g~\overline{v}~sg}~\Delta$ represents a cross-compartment call,
  where $\mathtt{f}$ is the name of the caller,
  $\mathtt{t}$ is the trace event produced by the call,
  $\mathtt{g}$ is the name of the callee,
  $\mathtt{\overline{v}}$ are the arguments,
  $\mathtt{sg}$ is the signature of the callee,
  and $\Delta$ are the memory deltas.
  \item $\mathtt{Ireturn~f~t~{v}}~\Delta$ represents a cross-compartment return,
  where $\mathtt{f}$ is the current procedure name,
  $\mathtt{t}$ is the trace event produced by the return,
  $\mathtt{v}$ is the return value,
  and $\Delta$ are 
  memory deltas.
  \item $\mathtt{Isys~f~t~ef~\overline{v}}~\Delta$ represents a system call,
  which is similar to the call case except that it requires the
  system call descriptor $\mathtt{ef}$
  instead of $\mathtt{g}$ and $\mathtt{sg}$.
\end{enumerate}

In this definition, \(\Delta\) represents a list of memory deltas.
A memory delta \(\delta\) records a sequence of all operations affecting the
contents of memory.
{}
For instance, \(\mathtt{delta\_store~ch~b~o~v~C}\)
represents a store at location \(\mathtt{(b, o)}\) of value \(\mathtt{v}\) performed
by compartment \(\mathtt{C}\), with memory chunk \(\mathtt{ch}\).
Given an informative event that contains a list of memory deltas \(\Delta\),
this records relevant operations since the last informative event.
Reapplying these allows us to reconstruct the global buffers in Clight.
%
We capture this intuition as part of a novel notion of well-formedness of informative
traces using an intermediate language.

Compared to \emph{data-flow events}~\cite{secureptrs} recording every
memory or register operation, our informative events only record operations
affecting global buffers.
While this means our back-translation doesn't support memory sharing, this also
considerably simplifies defining the back-translation and stating and
proving the invariants.

\paragraph{Well-formedness of informative events}
To characterize the informative traces that can be back-translated,
we define a state transition relation that abstracts over the \RISCV{} semantics.
%
States are triples containing parts of the \RISCV{} state at the points where
events are generated:
\(
\bstate \mathrel{:=} (\mathtt{f},\mathtt{M},\mathtt{fs}).
\)
Intuitively, the first element is the current procedure identifier;
the second element is the current memory in the \RISCV{} execution;
and the third element is a simplified view of the stack, seen as
a list of procedure identifiers.

The relation \(s \xrightarrow{\alpha} s'\) captures
the conditions for a state to take a step while producing an informative event, 
including relations among the global environment $\mathtt{ge}$ and procedures, memory deltas, and memory updates,
as well as the well-bracketedness of cross-compartment calls and returns.
It does not include information irrelevant to the back-translation to \clight{} though,
such as all the low-level details of the \RISCV{} semantics.
For instance, the following (simplified) rule describes the conditions for system calls:
\[
  \inferrule{
    \mathtt{mem\_delta\_apply~\bgenv~C~}\Delta~\mathtt{M_{1}} = \mathtt{M_{2}} \\
    \mathtt{globals\_scalar~\bgenv~C_{1}~M_{2}} \\
    \mathtt{wf\_deltas~}\delta
  }{
    \mathtt{(f_{1},M_{1},k)} \xrightarrow{\mathtt{Isys~f_{1}~(Event\_syscall~C~ef~\overline{v})~ef~\overline{v}~}\Delta}
  \mathtt{(f_1,M_{2},k)}
}
\]
This rule abstracts the corresponding rule in the \RISCV{} semantics:
it has the same conditions for the global environment, trace, shadow stack, and global buffers,
and adds conditions for updating the memory according to the memory deltas.
The rules for calls and returns are similar, except that they also update the stack.

As previewed in \autoref{sec:key-ideas:back-translation}, we say an informative
trace is well-formed when it is produced by the reflexive transitive closure of
this step relation, starting from a state corresponding to the initial state of \RISCV{}.
We prove that for any prefix $m$ of a trace produced by a \RISCV{} program,
there exists a well-formed informative trace $\btrace$, such that $\mathtt{proj}(\btrace) = m$.
Thanks to this result, we can forget about the \RISCV{} semantics, and simply
write a back-translation that takes as an input a well-formed informative trace, and then
prove its correctness using the well-formedness.

\paragraph{Back-translation function}\label{sec:bt-fun}
Our back-translation function constructs a \clight program by producing
procedures from the informative trace.
The back-translation converts each informative event to a \clight{} statement that produces the same event:
for instance, a call event is converted to a call instruction.
System calls are more interesting: before the system call, the
back-translation generates a list of instructions that write to each public variable according
to the last memory store recorded for that variable on the trace.
%
%
The stored values are guaranteed to be scalars by the well-formedness conditions,
which is necessary since \clight{} doesn't allow forging an arbitrary pointer.
Eventually, these events are wrapped inside a switch statement and a loop, as
described in \autoref{sec:key-ideas:back-translation}.

\paragraph{Correctness of the back-translation}
We prove a simulation between the intermediate language describing
the well-formedness of the informative trace and the \clight{} semantics of the program.
We prove that given two states, \(s_{i}\) of the intermediate language and \(s_{C}\) of \clight{}
that are related by our simulation invariant,
then a transition \(s_{i} \xrightarrow{\alpha} s_{i}'\) in the intermediate language
corresponds to a sequence of transitions \(s_{C} \xrightarrow{t}^{*} s_{C}'\) in \clight{},
where \(t = \mathtt{proj}(\alpha)\).
Additionally, we prove that \(s_{i}'\) and \(s_{C}'\) are related by our simulation invariant,
which does not require the memory of the intermediate language state to inject into the \clight{} memory; 
instead, we only maintain the invariant that the global buffers of each compartment are writable.
This invariant and the well-formedness of informative events provide enough information for us to
construct a memory injection for the global buffers at the point of a system call.
Specifically, when encountering a system call, we can use the facts stored in the rule 
above:
\(\mathtt{mem\_delta\_apply~\bgenv~C~}\Delta~\mathtt{M_{1}} = \mathtt{M_{2}}\),
\(\mathtt{wf\_delta}~\Delta\), and \(\mathtt{M_{2}}\) only contains scalars
(\(\mathtt{globals\_scalar~\bgenv~C_{1}~M_{2}}\)).
Using these we show that
the back-translation of this informative event writes all the necessary values
in the global buffers, and we can construct a memory injection that is
only defined on these buffers.
We can then apply the CompCert axiom discussed in the 2nd paragraph to prove
that the system call succeeds and produces the same event.

Compared to the data-flow back-translation~\cite{secureptrs}, we do not rely on
maintaining the invariant that the memories are related by a memory renaming (or a memory injection).
Instead, we only have to prove that the part of the memories dedicated to the global buffers
are related at specific points in the proof.
Establishing this relatedness is also made easier by the fact the global buffers only contain scalars,
which allows us to ignore the rest of the memory.

Because we use a finite, {\tt int64} counter to track how many events have been produced, 
we must limit the length of the (non-informative) trace prefix to $2^{64}$ events.
This limit on prefix lengths is, however, higher than the one coming from the
assumption that the compilation of the back-translated program succeeds (see
\autoref{sec:compiling-bt}).

\section{Recomposition Proof Details}\label{sec:recomposition}

\begin{figure*}
  \centering
\begin{subfigure}{0.3\textwidth}
  \centering
  \begin{tikzpicture}
    \tikzstyle{every node}=[BlueViolet]
    \tikzstyle{every path}=[BlueViolet]
    \node[draw,shape=circle,fill=BlueViolet,inner sep=0.5pt](ini1) {};
    \node[above left = 0.1pt of ini1, inner sep=0.5pt] (textini1) {\(s_{1}\)};
    \node[draw,shape=circle,fill=BlueViolet,inner sep=0.5pt,right = 2cm of ini1] (after1) {};
    \node[above right = 0.1pt of after1, inner sep=0.5pt] (textafter3) {\(s_{1}'\)};
    \node[draw,shape=circle,fill=BlueViolet,inner sep=0.5pt,below = of ini1] (ini2) {};
    \node[above right = 0.1pt of ini2, inner sep=0.5pt] (textini2) {\(s_{2}\)};
    \node[right = 2cm of ini2] (after2) {};
    \node[draw,shape=circle,fill=BlueViolet,inner sep=0.5pt,below = of ini2] (ini3) {};
    \node[below left = 0.1pt of ini3, inner sep=0.5pt] (textini3) {\(s_{3}\)};
    \node[draw,shape=circle,fill=PineGreen,PineGreen,inner sep=0.5pt,right = 2cm of ini3] (after3) {};
    \node[below right = 0.1pt of after3, inner sep=0.5pt] (textafter3) {\(\color{PineGreen}s_{3}'\)};

    \draw[->,shorten >= 2pt, shorten <= 2pt,thick] (ini1) to node [above] {\(\varepsilon\)} (after1.-180);
    \draw[->,shorten >= 2pt, shorten <= 2pt,color=PineGreen] (ini3) to node [above] {\(\color{PineGreen}\varepsilon\)} (after3.-180);

    \draw[shorten >= 2pt, shorten <= 2pt,thick] (ini1) to[bend right] node [left] {\(\equiv\)} (ini3);
    \draw[shorten >= 2pt, shorten <= 2pt, densely dashed,thick] (ini2) to[bend left] node [right] {\(\sim\)} (ini3);

    \draw[shorten >= 2pt, shorten <= 2pt, PineGreen] (after1) to[bend left] node [right] {\(\color{PineGreen}\equiv\)} (after3);
    \draw[shorten >= 2pt, shorten <= 2pt, densely dashed, PineGreen] (ini2) to[bend left] node [right] {\(\color{PineGreen}\sim\)} (after3);

    \draw[dotted,thick] (-0.75,0.5) rectangle (0.5,-2.5);
    \node[left = 0.75cm of ini3, inner sep=0.5pt] (m) {\(\mathcal{M}\)};

    \draw[dotted,thick,PineGreen] (1.75,0.5) rectangle (3,-2.5);
    \node[right = 0.92cm of after3, inner sep=0.5pt] (mm) {\(\color{PineGreen}\mathcal{M}\)};
  \end{tikzpicture}
  \vspace{-0.5em}
  \caption{Silent step in strongly related states}\label{fig:recomposition-diagrams:silent-strong}
\end{subfigure}
\begin{subfigure}{0.3\textwidth}
  \centering
  \begin{tikzpicture}
    \tikzstyle{every node}=[BlueViolet]
    \tikzstyle{every path}=[BlueViolet]

    \node[draw,shape=circle,fill=BlueViolet,inner sep=0.5pt](ini1) {};
    \node[above left = 0.1pt of ini1, inner sep=0.5pt] (textini1) {\(s_{1}\)};

    \node[draw,shape=circle,fill=BlueViolet,inner sep=0.5pt,below = of ini1](ini2) {};
    \node[above right = 0.1pt of ini2, inner sep=0.5pt] (textini2) {\(s_{2}\)};

    \node[draw,shape=circle,fill=BlueViolet,inner sep=0.5pt,right = 2cm of ini2] (after2) {};
    \node[above right = 0.1pt of after2, inner sep=0.5pt] (textafter1) {\(s_{2}'\)};

    \node[draw,shape=circle,fill=BlueViolet,inner sep=0.5pt,below = of ini2] (ini3) {};
    \node[below left = 0.1pt of ini3, inner sep=0.5pt] (textini3) {\(s_{3}\)};

    \node[draw,shape=circle,fill=BlueViolet,inner sep=0.5pt,right = 2cm of ini1] (equ1) {};
    \node[above right = 0.1pt of after1, inner sep=0.5pt] (textafter3) {\(s_{1}\)};
    \draw[double,shorten >= 2pt, shorten <= 2pt] (ini1) to[bend left] node [below] {\(=\)} (equ1);

    \node[draw,shape=circle,fill=PineGreen,inner sep=0.5pt,right = 2cm of ini3] (equ3) {};
    \node[below right = 0.1pt of after3, inner sep=0.5pt] (textafter3) {\(s_{3}\)};
    \draw[double,shorten >= 2pt, shorten <= 2pt] (ini3) to[bend right] node [above] {\(=\)} (equ3);

    \draw[->,shorten >= 2pt, shorten <= 2pt,thick] (ini2) to node [above] {\(\varepsilon\)} (after2.-180);

    \draw[shorten >= 2pt, shorten <= 2pt,thick] (ini1) to[bend right] node [left] {\(\equiv\)} (ini3);
    \draw[shorten >= 2pt, shorten <= 2pt, densely dashed,thick] (ini2) to[bend left] node [right] {\(\sim\)} (ini3);

    \draw[shorten >= 2pt, shorten <= 2pt, densely dashed, PineGreen] (ini3) to[bend right] node [right] {\(\color{PineGreen}\sim\)} (after2);

    \draw[dotted,thick] (-0.75,0.5) rectangle (0.5,-2.5);
    \node[left = 0.75cm of ini3, inner sep=0.5pt] (m) {\(\mathcal{M}\)};

    \draw[dotted,thick,PineGreen] (1.75,0.5) rectangle (3,-2.5);
    \node[right = 0.92cm of after3, inner sep=0.5pt] (mm) {\(\color{PineGreen}\mathcal{M}\)};
  \end{tikzpicture}
  \vspace{-0.5em}
  \caption{Silent step in weakly related states}\label{fig:recomposition-diagrams:silent-weak}
\end{subfigure}
\begin{subfigure}{0.3\textwidth}
  \centering
  \begin{tikzpicture}
    \tikzstyle{every node}=[BlueViolet]
    \tikzstyle{every path}=[BlueViolet]

    \node[draw,shape=circle,fill=BlueViolet,inner sep=0.5pt](ini1) {};
    \node[above left = 0.1pt of ini1, inner sep=0.5pt] (textini1) {\(s_{1}\)};

    \node[draw,shape=circle,fill=BlueViolet,inner sep=0.5pt,right = 2cm of ini1] (after1) {};
    \node[above right = 0.1pt of after1, inner sep=0.5pt] (textafter1) {\(s_{1}'\)};

    \node[draw,shape=circle,fill=BlueViolet,inner sep=0.5pt,below = of ini1] (ini2) {};
    \node[above right = 0.1pt of ini2, inner sep=0.5pt] (textini2) {\(s_{2}\)};

    \node[draw,shape=circle,fill=BlueViolet,inner sep=0.5pt,right = 2cm of ini2] (after2) {};
    \node[above right = 0.1pt of after2, inner sep=0.5pt] (textafter2) {\(s_{2}'\)};

    \node[draw,shape=circle,fill=BlueViolet,inner sep=0.5pt,below = of ini2] (ini3) {};
    \node[below left = 0.1pt of ini3, inner sep=0.5pt] (textini3) {\(s_{3}\)};
    \node[draw,shape=circle,fill=PineGreen,PineGreen,inner sep=0.5pt,right = 2cm of ini3] (after3) {};
    \node[below right = 0.1pt of after3, inner sep=0.5pt] (textafter3) {\(\color{PineGreen}s_{3}'\)};

    \draw[shorten >= 2pt, shorten <= 2pt,thick] (ini1) to[bend right] node [left] {\(\equiv\)} (ini3);
    \draw[shorten >= 2pt, shorten <= 2pt, densely dashed,thick] (ini2) to[bend left] node [right] {\(\sim\)} (ini3);

    \draw[shorten >= 2pt, shorten <= 2pt, densely dashed, PineGreen] (after1) to[bend left] node [right] {\(\color{PineGreen}\sim\)} (after3);
    \draw[shorten >= 2pt, shorten <= 2pt, PineGreen] (after2) to[bend right] node [left] {\(\color{PineGreen}\equiv\)} (after3);

    \draw[->,shorten >= 2pt, shorten <= 2pt,thick] (ini1) to node [above] {\(a\)} (after1.-180);
    \draw[->,shorten >= 2pt, shorten <= 2pt,thick] (ini2) to node [above] {\(a\)} (after2.-180);
    \draw[->,shorten >= 2pt, shorten <= 2pt,color=PineGreen] (ini3) to node [above] {\(\color{PineGreen}a\)} (after3.-180);

    \draw[dotted,thick] (-0.75,0.5) rectangle (0.5,-2.5);
    \node[left = 0.75cm of ini3, inner sep=0.5pt] (m) {\(\mathcal{M}\)};

    \draw[dotted, PineGreen] (1.75,0.5) rectangle (3,-2.5);
    \node[right = 0.92cm of after3, inner sep=0.5pt] (mm) {\(\color{PineGreen}\mathcal{M}\)};
  \end{tikzpicture}
  \vspace{-0.5em}
  \caption{Non-silent step with swapping relations}\label{fig:recomposition-diagrams:event-swap}
\end{subfigure}
  \vspace{-1em}
  \caption{Recomposition diagrams}\label{fig:recomposition-diagrams}
  \vspace{-1em}
\end{figure*}
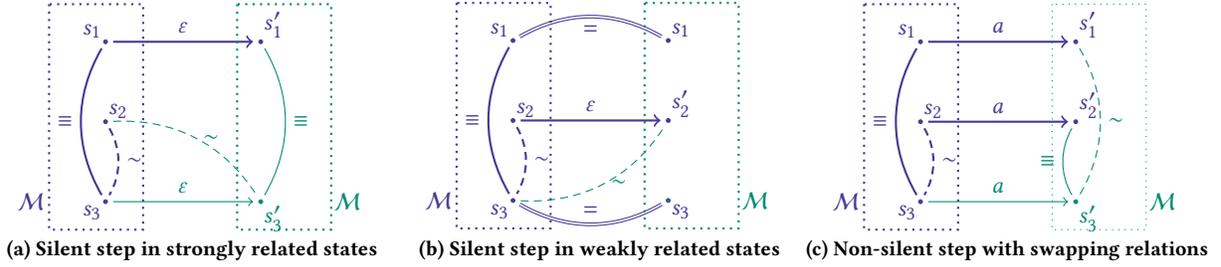

In \autoref{sec:key-ideas:recomposition}, we described our recomposition proof
technique at a high level.
Here we detail the proof diagrams that, together, imply the three-way
simulation of recomposition.
The most important three of these diagrams are depicted in \autoref{fig:recomposition-diagrams}.
As in \autoref{sec:key-ideas:recomposition}, in each diagram, each row represents one of the 3 executions.
Arrows represent execution steps, and are annotated with either an event \(a\) or silence \(\varepsilon\).
We use thick, dark purple for the assumptions, and dark green for the
conclusions we have to prove (including existence of states).
%
%
We depict in dashed line the weak relation, and in plain line the strong relation;
and we use dotted rectangles to depict the mixed relation.
We denote equality via a double line in \autoref{fig:recomposition-diagrams:silent-weak}.
\jt{Is it worth repeating the explanation of how to read the diagram here?}\ch{Let's see how bad we do on space :)}
\jt{What we could write to save space:
  The diagrams read similarly to those of \autoref{sec:key-ideas:recomposition}.
  In addition, we depict in dashed line the weak relation, and in plain line the strong relation;
  and we use dotted rectangles to depict the mixed relation.}

\autoref{fig:recomposition-diagrams:silent-strong} describes the case where \(s_{1}\)
and \(s_{3}\) are taking a silent step synchronously and can be read as follows: starting
from three states \(s_{1}\), \(s_{2}\), and \(s_{3}\) related by \(\mathcal{M}\),
such that \(s_{1} \equiv s_{3}\) and \(s_{2} \sim s_{3}\), and such that \(s_{1}\) steps
silently to \(s_{1}'\), we have to prove that \(s_{3}\) also steps silently to another
state \(s_{3}'\), that the weak and strong relation are reestablished for \(s_{3}'\)
(\IE \(s_{1}' \equiv s_{3}'\) and \(s_{2} \sim s_{3}'\)),
and that the mixed relation is also reestablished.
Similarly, \autoref{fig:recomposition-diagrams:silent-weak} describes the case where \(s_{2}\)
takes a silent step; because it is only weakly related to \(s_{3}\), we do not
require either \(s_{1}\) or \(s_{3}\) to take steps as well, but we must still reestablish
the relations.
Finally, \autoref{fig:recomposition-diagrams:event-swap}
describe the case where both executions produce an event \(a\).
Given \(s_{1} \equiv s_{3}\) and \(s_{2} \sim s_{3}\), related by \(\mathcal{M}\), such that
\(s_{1} \xrightarrow{a} s_{1}'\) and \(s_{2} \xrightarrow{a} s_{2}'\), one must
prove the existence of \(s_{3}'\) such that \(s_{3} \xrightarrow{a} s_{3}'\), \IE that
the three executions advance in lockstep.
If the event constitutes a change of control between the two sides (program and context)
weak and strong relations are be swapped as illustrated in the diagram; but we also
have a similar diagram where the relations are not swapped.

We prove our 8 diagrams together imply the existence of the above three-way simulation, and
hence imply the recomposition theorem.
To do so, we simply instantiate \(\mathcal{R}\) with the conjunction
of weak, strong, and mixed relations applied to the appropriate cases.

\paragraph{Applying diagrams to our setting}
We now explain how we instantiate the parameters of our proof diagrams.
At the \RISCV{} level, states are the disjoint union of regular states \(s {=} (\ii{regs}, \mathtt{M}, \ii{st})\)
and return-states \(\ii{rs} {=} (\ii{regs}, \mathtt{M}, \ii{st}, \mathtt{C})\) where \(\ii{regs}\) is a register set,
\(\mathtt{M}\) is a memory, \(\ii{st}\) a shadow stack, and \(\mathtt{C}\) a compartment name
recording which compartment the execution is returning from.
To handle the possibility of having different allocation behavior in each execution, we relate memories
and values using two memory injections \(j_{1}\) and \(j_{2}\), one for each of the original executions.
These memory injections are only defined on their execution's kept compartments, and do not describe
the other compartments' memory.
These memory injections parameterize the relations,
are kept as part of the ghost state we maintain, and
are updated during the execution.
As described previously, the weak and the strong relations relate the memories of the run
according to these injections, the strong relation also relates the registers,
and the mixed relation relates the stacks, requiring that the stacks represent
the same series of calls with the same arguments.

Only the proofs of the 3 diagrams from \autoref{fig:recomposition-diagrams} are
interesting, since the remaining 5 follow by exploiting symmetries of the
diagrams and of the relations.
The first two diagrams require that silent steps don't affect other compartments.
This wouldn't hold if stack frames storing spilled arguments were still writable
by the caller, since otherwise a compartment could surreptitiously communicate
by changing previously passed arguments in an old stack frame.
We discovered this issue while trying to complete the proof of these two
diagrams and failing, as we couldn't prove that a compartment reading an
argument from the stack would get related values according to the memory
injection of its side (it could instead get a pointer from the other side that's
not in the memory injection).
This required us to step back and change the semantics of \RISCV{} assembly to
make old arguments passed on the stack read only, as described in
\autoref{sec:extending-compcert}.
Finally, the last diagram, where an event is produced, is proved by
re-establishing the three relations after producing the correct event.
This is made possible by the fact that all information that is shared between
the two compartments on a call or return appears in the trace event.

While our proof structure is similar to the turn-taking simulation of El-Korashy~\ETAL~\cite{secureptrs},
we greatly benefit from the proof diagrams above, as they give us
a clear structure to organize the proof.
Moreover, because we rely on CompCert's memory injections instead of ad-hoc
memory renamings, we are able to reuse much of the machinery that already exists
for the compiler correctness proof.

\section{Compiling back-translation result}
\label{sec:compiling-bt}

The FCC statement of Abate~\ETAL~\cite{AbateABEFHLPST18} (reproduced in
\autoref{def:fcc}) assumed that the compiler is a total function.
While this was true of their very simple compiler, it is not the case for
realistic compilers like CompCert, so our FCC theorem needs 
an extra assumption that the compiler can successfully compile $\src{C}$ and $\src{P}$:

\begin{theorem}[Forward Compiler Correctness (FCC)]\label{thm:fcc}
$\forall \src{C}~\src{P}.$
\[
  \forall m {\not\ni} \undefi.~
  (\src{C} \linksrc \src{P}) \semsrc m \land \cmp{\src{C}} \text{and } \cmp{\src{P}} \text{defined}
  \Rightarrow (\cmp{\src{C}} \linktrg~ \cmp{\src{P}}) \semtrg m.
\]
\end{theorem}

This extra assumption in FCC leads to a new assumption in our \rscdcmd proof, namely
that the result of our
back-translation can be successfully compiled.
But this is not at all easy to prove:
CompCert has many sources of partiality and it is not feasible to guarantee in
advance that a (well-typed) C program will be successfully compiled.
Two passes in CompCert, register allocation and linearization, are not verified
but rely on translation validation, which can fail.
The data-flow analyzer used in several optimization passes can fail if
the analysis doesn't converge after a very large number of steps (e.g. $10^{12}$).
Several passes (notably Asmgen) use errors to rule out ill-formed code that
should never have made it this far, but which is easier to recheck than to prove
impossible.
Finally, several passes (CminorGen, Inlining, Stacking, \ETC) put constraints on
the size of the generated stack frames to ensure that offsets within these
frames don't overflow a machine word.

Compiling the back-translation is
just an artifact of the
high-level proof structure (\autoref{fig:rsc-dc-md-proof}), which uses
compiler correctness for whole programs to repeatedly move between the source
and target languages.
Therefore we assume as an axiom that the result of our back-translation,
on any trace prefix below a certain length, \emph{can} be successfully compiled.
The length bound is needed to account for machine
words being finite and other such finite resources.

\begin{assumption}[Back-translation successfully compiles]\label{ass:bt-compiles}
\[
\begin{array}{l}
\forall K.~ 
\forall I.~ 
\forall \trg{W_T} {:} I.~
\forall m \not\ni \undefi.~ |m| \leq \ii{MAX\_TRACE\_LENGTH} \Rightarrow\\
  \forall \trg{p} : \trg{W_T} \semtrg m.~
  \forall \src{C} {:} \lfloor I \rfloor_{\ii{K}}.~
  \src{C} {=} \underset{k\in K}{\biglinksrc}\back{(I,\trg{p},k)} \Rightarrow
  \cmp{\src{C}} \text{is defined}
\end{array}
\]
\end{assumption}

The next paragraphs report how we have systematically tested this assumption for
a large number of trace prefixes.
In the future one could envision using more compositional compiler
correctness results than that of CompCert, recent~\cite{SongCKKKH20, KoenigS21,
  SammlerSSDKGD23, ZhangWWKS24} or upcoming, to potentially overcome the need
for this assumption.
For now though, we take this assumption as a reasonable cost to pay for a
secure compilation proof technique that is the first to scale up to a
realistic compiler like CompCert and that only
requires an operational semantics for whole programs, which is not
compositional, but which simplifies proofs, including CompCert's existing
compiler correctness proof that we extended here to isolated compartments.


\paragraph{Property-based testing of \autoref{ass:bt-compiles}}
%
We systematically tested that the Clight programs constructed by our back-translation function can be compiled with CompCert again.
Concretely, we experimentally test that $\mathtt{ccomp}(\mathtt{bt\_fun}(bt, env))$ succeeds for
random but \textit{consistent} informative traces $bt$ and environments $env$ (\autoref{sec:back-translation}). 
The environment defines a set of compartments, their interfaces, and the available procedures that can
be referenced in the traces. We generate random environments by deriving them from random, undirected
and connected graphs $\mathcal{G} = (V, E)$. Each vertex $v \in V$ represents a compartment and we associate
it with a random, non-empty set of procedures and signatures $\mathtt{v.exports}$. Further, for each $(u, v) \in E$ we set
$\mathtt{u.imports}$ to a random, non-empty subset of $\mathtt{v.exports}$ and vice-versa (c.f. \autoref{sec:extending-compcert}).
The trace is generated 
consistent with the environment such that (1) each procedure call is allowed
(c.f.~ \autoref{sec:extending-compcert}); (2) two calls to the same procedure use the same signature and (3) all values passed as arguments
or return values match the signature. For efficiency, we only generate values and in particular memory deltas that
are explicitly inspected in the back-translation function and not trivially compiled to skips.

We have been able to successfully compile all Clight programs produced by
back-translation for more than 100k pairs of generated environments and traces
with up to 880 events and close to 400 events on average.
Individual tests with significantly longer traces of more than 150k events also succeeded,
but the growing computational costs make it hard to test even longer traces.
In total, the traces contained over 16M $\mathtt{Icall}$ and $\mathtt{Ireturn}$, 7M
$\mathtt{Isys}$ and 500M $\mathtt{delta\_storev}$ instances.

\section{Top-level \rscdcmd theorem}
\label{sec:top-level-theorem}


We follow the general proof diagram from \autoref{fig:rsc-dc-md-proof} to
assemble our previous theorems and obtain the final result that our compilation
chain satisfies a variant of \rscdcmd{} from \autoref{def:rsc-dc-md}.
%

\begin{theorem}[\rscdcmd{}]\label{thm:rscdcmd}
SECOMP satisfies \rscdcmd{} for all trace prefixes $m$ such that $|m| \leq \ii{MAX\_TRACE\_LENGTH}$.
\end{theorem}

The size of the prefixes supported by this theorem is restricted
by the need to successfully compile
the results of the back-translation (\autoref{ass:bt-compiles}), which we have
systematically tested as described in \autoref{sec:compiling-bt}. 
A second disclaimer is that we have not yet finished integrating the Coq proof
of this theorem with the Coq proofs of the individual steps,
and as mentioned
in \autoref{sec:intro}, the proofs of back-translation,
and blame are still on separate branches.

\section{Enforcement using capabilities}
\label{sec:caps}

To show that the capability abstraction we added to the semantics of CompCert's
\RISCV assembly is practically implementable at a lower level, we designed a
capability backend for SECOMP.
The backend targets an extension of the CHERI \RISCV
architecture~\cite{CHERI-ISA}, which provides hardware capabilities; \IE
unforgeable pointers with base and bounds that cannot be circumvented.
%
%
While various secure calling conventions targeting capabilities have been
proposed in recent years~\cite{GeorgesGSTTHDB21, SkorstengaardDB21,
SkorstengaardDB20, TsampasDP19}, our backend is based on the most recent
proposal of Georges~\ETAL~\cite{GeorgesTB22}, which uses not only the standard
capabilities described above, but also CHERI's local
capabilities~\cite{SkorstengaardDB20}, entry, and sealed capabilities.
%
In particular, stack pointers are implemented as local capabilities which can only be stored
on the stack or in registers.
Hence, a compartment cannot save a capability to some part of the stack for later use, and
cannot modify the value of the arguments it passed \textit{a posteriori}, an attack we
prevent for languages higher in the compilation chain by making some stack
frames read-only (see \autoref{sec:extending-compcert}).
Additionally, this calling convention is based on two newly proposed kinds of capabilities:
uninitialized~\cite{GeorgesGSTTHDB21} and directed~\cite{GeorgesTB22}.
%
In short, uninitialized capabilities prevent reading old values from the stack
without excessive clearing~\cite{GeorgesGSTTHDB21}, and directed capabilities
support efficient implementation of stack safety~\cite{GeorgesTB22}.
Our backend targets a lower-level variant of CompCert's \RISCV assembly language
with a flat memory model and extended with all these capabilities.

While our calling convention is inspired by Georges~\ETAL~\cite{GeorgesTB22} we
had to adapt that design to our setting in two ways: First, because we only
enforce compartment isolation, not memory safety, we represent pointers as
offsets into a large stack capability or into per-compartment heap capabilities.
By not using directed capabilities for stack pointers, we overcome a potential
limitation of Georges~\ETAL's~\cite{GeorgesTB22} calling convention and can
store cyclic data structures on the stack.
Second, compared to Georges~\ETAL~\cite{GeorgesTB22} we consider a stronger
attacker model, in which both the caller and the callee compartments of a call
can be compromised.
In our model we thus need to always maintain the distinction between the caller
and callee compartments and enforce that no capabilities are exchanged between the two.
We achieve this by adding privileged wrappers for calls and returns, which ensure
that the passed arguments/returns are not capabilities, and which clear all remaining registers.

We built a prototype implementation of this backend in Coq that can already compile
simple examples, but that has not yet been thoroughly tested and is not verified.
%
In the short run, one can use property-based testing to get more confidence
in its correctness and security.
We are also considering the design of a second capability backend inspired by
the original work of Watson~\ETAL~\cite{WatsonWNMACDDGL15} and implemented in
CheriBSD~\citeFull{CheriBSDComp}{CheriBSD}, which only uses the existing features of CHERI.
This second design, however, requires a split stack layout, which is allowed by the
C standard and the CompCert memory model, but which changes the \RISCV calling convention.
In the long run, formally verifying such backends is a very interesting
research challenge, as also discussed in \autoref{sec:future}.

\section{Related work}\label{sec:related}

As explained in \autoref{sec:background}, we directly build on the work of
Abate~\ETAL~\cite{AbateABEFHLPST18}, in particular by reusing their \rscdcmd
secure compilation criterion and their high-level proof structure.
Scaling up these ideas from a very simple compiler for a toy programming
language all the way to a verified compiler for the realistic C language was an
open research challenge that we overcome in this paper.
In our realistic setting, the back-translation, recomposition, and blame steps
are more interesting and require several proof engineering novelties and also
more sophisticated invariants involving memory injections.
Moreover, for the compiler correctness steps, which were assumed by
Abate~\ETAL~\cite{AbateABEFHLPST18}, we show that with careful design we can
extend the massive CompCert proof to compartments with a manageable amount of
effort (only around 9\% size increase).

The security proofs of Abate~\ETAL~\cite{AbateABEFHLPST18} and also a later variant
with pointer passing~\cite{secureptrs} (also discussed in \autoref{sec:future})
are both mechanized in Coq.
Even if compiler correctness is assumed,
these are among the few proofs of secure compilation against adversarial
contexts (i.e., for criteria like full abstraction and RSP~\cite{AbateBGHPT19})
that have been mechanized in a proof assistant, with the majority of work in
this space being proved only on paper, usually for even simpler languages and
compilers~\citeFull{AgtenSJP12, JuglaretHAEP16, PatrignaniC15, PatrignaniASJCP15, 
  MarcosSurvey, AbateBGHPT19, PatrignaniG21, BusiNBGDMP21,
  akram-capabileptrs}{AbadiFG02, AbadiP12, AhmedB08, AhmedB11, FournetSCDSL13,
  NewBA16,JagadeesanPRR11, PatrignaniDP16, PatrignaniG17}.
One proof that was fully mechanized in Coq is that of
Devriese~\ETAL~\cite{DevriesePPK17}, who prove modular full abstraction by
approximate back-translation for a compiler from the simply typed to the
untyped $\lambda$-calculus.
Jacobs~\ETAL~\cite{JacobsDT22} prove in Coq the purity of a Haskell-like ST monad
stated as full abstraction of a translation from a pure language.
Georges~\ETAL~\cite{GeorgesTB22} prove in Coq the security of their calling
convention (a variant of which we also use in \autoref{sec:caps}) stated as the
full abstraction of the identity compiler between a secure overlay semantics and
the actual semantics of a simple idealized assembly language.
Finally, Abate~\ETAL~\cite[\S 7.1]{difftraces} verify a very simple compiler in
Coq illustrating secure compilation when the target language has additional
trace events that are not possible in the source.

A more realistic related work is
CompCertSFI~\cite{BessonBDJW19}, which builds on previous ideas by
Kroll~\ETAL~\cite{KrollSA14} to implement portable SFI as a source-to-source
transformation in Cminor, an intermediate language of CompCert that comes before
optimizations.
Pointers are represented as integers and masked in order to offset into a single
big array representing all of the sandbox's memory.
In addition to masking pointers and using trampolines for functions pointers,
CompCertSFI instruments the program to prevent any undefined behavior.
This is needed to properly preserve the main security result, showing that all
memory accesses stay within the sandbox, down to CompCert's assembly language.
An experimental evaluation shows that the overhead of CompCertSFI comes mostly
from CompCert itself performing less aggressive optimizations than GCC and Clang.
When the proposed SFI transformation is instead used with GCC or Clang the
overheads are generally competitive with (P)NaCl~\cite{YeeSDCMOONF10}.
By implementing SFI in an early intermediate language, CompCertSFI can take
advantage of all the compiler's optimizations as well as the alignment
analysis added by the authors.

Our implementation strategy is different and is not targeted specifically at
SFI, but instead at being able to take advantage of hardware features for
compartment isolation such as
 capabilities~\citeFull{BauereissCSAESB22, CHERI-ISA}{AmarCCFLLNMTWX23}
or the recently proposed support for Hardware Fault
Isolation~\cite{NarayanGTRM0FVL23}.
%
%
Another difference is that SECOMP supports an arbitrary number of
mutually-distrustful compartments that can interact by calling each other
according to clearly specified interfaces.
Finally, while one could potentially extend CompCertSFI to achieve a security
notion similar to the original RSP~\citeFull{AbateBGHPT19, difftraces,
  PatrignaniG21}{PatrignaniG17} (so without mutual distrust), this would
require more work, for instance proving compiler correctness with respect to the
semantics of source programs, bridging the gap between the memory model used by
CompCert and the single memory block model used by CompCertSFI, and devising
a verified back-translation between Cminor and higher CompCert languages like C
or Clight.



Another verified SFI compiler is vWasm~\cite{BosamiyaLP22}, for which the
authors proved in \fstar{}~\cite{fstar-popl2016} that Wasm code compiled to x86
can only interact with its host environment via an explicitly provided API.
While this security guarantee and interaction model is similar to that of
CompCertSFI, the vWasm implementation doesn't take advantage of all standard
compiler optimizations, which leads to some performance loss.
The security guarantee only talks about the x86 semantics and, as opposed to our
work, does not aim at providing source-level security reasoning, even at the Wasm level.

Another realistic work in this space is that of Derakhshan~\ETAL~\cite{DerakhshanZVJ23}, who
devise a methodology to break up Trusted Execution Environment (TEE) software
into concurrently executing C compartments \ifallcites(überobjects~\cite{VasudevanCMJD16})\fi{}
whose security is compositionally verified using semi-automatic tools
\ifallcites(Frama-C~\cite{BaudinBBCKKMPPS21})\fi{} and which are then correctly compiled using a
verified compiler \ifallcites(CASCompCert~\cite{JiangLXZF19})\fi.
The formalization of this work is done on paper and the main assumption is that
all C compartments are verified, which seems realistic only for small,
highly privileged pieces of code, like the TEEs this work considers.
Our focus is instead on machine-checked proofs and on compartmentalized C code
that can't be formally verified to be even free of undefined behaviors.


Our work targets a variant of RSP, but such preservation of property classes against
adversarial contexts is not the only kind of formally secure compilation.
Another important kind aims at preserving specific noninterference properties
against passive side-channel attackers.
For instance, preservation of cryptographic constant time was proved for both
the CompCert~\cite{BartheBGHLPT20} and Jasmin~\citeFull{BartheGLP21}{AlmeidaBBGKL0S20}
verified compilers.
Another example is guaranteeing that protection against memory probing is
preserved by CompCert~\cite{BessonDJ19}.

Other formal verification work looks at security of low-level enforcement
mechanisms, without involving a compiler from a higher-level language.
For instance, SFI mechanisms for both x86~\cite{MorrisettTTTG12} and
ARM~\cite{ZhaoLSR11} were proved correct in a proof assistant with respect to
the semantics of these complex architectures.
In these works communication between low-level compartments is done by jumping
to a specified set of entry points, while we consider a more structured model
that also enforces the correct return discipline.
Other work in this space looks at the basic security properties of capability
machines, from simpler ones~\cite{StrydonckGGTTPB22, HuyghebaertKRD23} to more
realistic ones like CHERI~\cite{NienhuisJBFR0NN20} and Arm
Morello~\cite{BauereissCSAESB22}.




\section{Future Work}\label{sec:future}

\paragraph{Compiling realistic compartmentalized applications}
At the moment we have evaluated SECOMP only on very simple code, for
instance C variants of the examples of Abate~\ETAL~\cite{AbateABEFHLPST18}.
The main obstacle to compiling more realistic compartmentalized applications is
the current inability to communicate non-scalar data.
An immediate solution would be to add an IPC-like mechanism for passing the
contents of buffers between compartments.
A more ambitious solution (discussed next) would be to allow sharing
memory by passing capabilities on a machine like CHERI~\cite{CHERI-ISA}.
Another way to make SECOMP more practical would be to extend our interfaces with
more fine-grained access control policies for
IO~\cite{abadi2003access\ifanon\else, AndriciCHMRTW24\fi}.
Finally, another interesting direction is connecting with tools for
semi-automated compartmentalization of realistic C applications~\cite{GudkaWACDLMNR15}.

\paragraph{Pointer passing and memory sharing}
As with mainstream compartment isolation mechanisms (\EG SFI or OS processes),
we assume that compartments can only communicate via scalar values, but cannot
pass each other pointers to share memory.
While secure pointer passing is possible to implement
efficiently on a capability machine like CHERI~\cite{CHERI-ISA} or on the micro-policies tagged
architecture~\cite{micropolicies2015} and this would allow a more efficient
interaction model that is also natural for C programmers, the main challenge one still
has to overcome is {\em proving} secure compilation at scale in the presence of
such fine-grained, dynamic memory sharing.

Recent work by El-Korashy~\ETAL~\cite{secureptrs} in a much simpler setting
shows that it is indeed possible to prove in Coq
the security of an extension of Abate~\ETAL's~\cite{AbateABEFHLPST18}
compiler that allows passing secure pointers (\EG capabilities) between compartments.
With such fine-grained memory sharing, however, proofs become more
challenging and the proof technique of
El-Korashy~\ETAL{} led to much larger proofs
and still has conceptual limitations that one
would need to overcome for it to work for CompCert, in particular for supporting
memory injections.
In fact, even extending CompCert's compiler correctness proof to passing
arbitrary pointers seems a challenge, since it would imply a significant change
to CompCert's trace model.
In the nearer future we will try to allow more limited forms of memory
sharing between compartments, for instance of statically allocated buffers,
which could be passed without significantly changing CompCert's trace model.

\paragraph{Building and verifying lower-level backends}
%
%
Like CompCert's correctness proofs, the SECOMP security proofs currently stop at
CompCert's \RISCV assembly language.
We extended this language with
the abstraction of isolated compartments, which formally defines {\em what}
compartment isolation enforcement should do, but which leaves the {\em how} to
lower-level enforcement mechanisms.
%
Beyond the capability-based backend of \autoref{sec:caps},
various other enforcement mechanisms
should be possible, including SFI~\citeFull{sfi_sosp1993, YeeSDCMOONF10,
Tan17}{KolosickNJWLGJS22, JohnsonLZGNSSB23} and tagged
architectures~\cite{micropolicies2015, pump_asplos2015}, as shown in a much
simpler setting by Abate~\ETAL~\cite{AbateABEFHLPST18}.
Moreover, WebAssembly components~\citeFull{WasmCompWAW, WasmCompDoc,
  HaasRSTHGWZB17, BosamiyaLP22}{KolosickNJWLGJS22, WattRPBG21} could also be a
target for such a backend.

At the moment the existing lower-level backends are all unverified.
%
Extending the secure compilation proofs down to cover them is a formidable
research challenge that we leave for future work.
All existing secure compilation proof techniques in this
space~\cite{AbateBGHPT19,MarcosSurvey}, including the one we use in the current
paper~\cite{AbateABEFHLPST18}, have their origin in proof techniques for full
abstraction~\cite{MarcosSurvey}.
But once the memory layout becomes concrete~\cite{WangWS19,WangXWS20} and the
code is explicitly stored in memory, we can no longer hide all
information about the compartments' code, as would be needed for full
abstraction (or in our case for recomposition), so new proof techniques will
be needed.



\paragraph{Targeting other hardware architectures}
While SECOMP currently targets RISC-V for simplicity, the biggest part of
CompCert is architecture independent, and our extension preserves this feature
and the only architecture-specific pass is still between Mach and assembly.
Also extending the CompCert passes from Mach to x86 or Arm assembly seems
feasible, and in particular adding the compartment abstraction to CompCert's
semantics for x86 and Arm would be very similar to what we already did for
RISC-V. A bigger challenge would be reproving recomposition: while some parts of
our proof are generic, such as reducing recomposition to the diagrams from
\autoref{fig:recomposition-diagrams}, proving these diagrams for huge
instruction sets like x86 and Arm would be very tedious.
One idea to overcome this challenge would be to assume that an attacker can only
run instructions produced by our compiler, and to enforce this in lower-level
backends using a combination of W\^{}X memory protection and some amount of
control-flow integrity.

Even on these architectures, lower-level backends could implement the
compartment abstraction in various ways. On a capability machine like Arm
Morello~\cite{BauereissCSAESB22}, one could still do hardware-supported
enforcement using capabilities, as outlined at the end of \autoref{sec:caps}.
On a modern x86 one could potentially make use of MPK for
gaining efficiency~\cite{Vahldiek-Oberwagner19}.
And if everything else fails, one could still enforce compartment isolation in
software using SFI.

\paragraph{From safety to hypersafety}
Another interesting direction is extending SECOMP to stronger
criteria beyond robust preservation of safety, in particular to
hypersafety~\cite{AbateBGHPT19}, such as data confidentiality.
We expect that SECOMP can be easily adapted to these stronger criteria, by for
instance always clearing registers before changing compartments, and also that
our RSP proof technique can still apply, by only extending the
back-translation step to take finite sets of trace prefixes as
input\ifanon~\cite{AbateBGHPT19}\else~\cite{AbateBGHPT19,nanopass}\fi.
A very interesting and challenging future work
is actually enforcing robust preservation of hypersafety in the
lower-level backends with respect to side-channel attacks, including
devastating micro-architectural attacks like Spectre~\cite{PatrignaniGu21}.


\paragraph{Compositional compiler correctness}
Reusing the massive correctness proof of CompCert was definitely worth it for
our work, yet the limited compositionality of CompCert~\cite{KangKHDV15} lead to
a proof technique that relies on an extra assumption (\autoref{ass:bt-compiles})
that realistically we could only test.
As mentioned in \autoref{sec:compiling-bt}, more compositional compiler
correctness results, recent~\cite{SongCKKKH20, KoenigS21, SammlerSSDKGD23,
  ZhangWWKS24} or upcoming, could potentially remove the need for this assumption.
Moreover, such compositional compiler correctness results could potentially make
our architecture-specific proofs easier, since recomposition could be split into
a decomposition step for the target and a composition step for the
source~\citeFull{AbateBGHPT19, JuglaretHAEP16, PatrignaniC15, PatrignaniASJCP15,
akram-capabileptrs, PatrignaniG21}{PatrignaniG17}.

\paragraph{Dynamic compartment creation and dynamic privileges}
SECOMP uses a static notion of compartments and static interfaces to restrict
their privileges.
SECOMP compartments are defined statically by the source program, so are a form of
code-based compartmentalization.
In the future one could also explore dynamic compartment creation, which would
allow for data-based compartmentalization~\cite{GudkaWACDLMNR15}, \EG one
compartment per incoming network connection or one compartment per web browser
tab or plugin~\cite{ReisG09}.
It would also be interesting to investigate dynamic privileges for compartments,
\EG dynamically sharing memory by passing secure pointers (as discussed above),
dynamically changing the compartment interfaces~\cite{seL4:Oakland2013}, or
history-based access control~\cite{abadi2003access\ifanon\else, AndriciCHMRTW24\fi}.



\ifanon\else
\paragraph{Acknowledgments}
We thank Adrien Durier for participating in early discussions about this work.
We are also grateful to the anonymous reviewers at PriSC'23 and CCS'24 for their helpful feedback.
This work was in part supported
by the \grantsponsor{1}{European Research Council}{https://erc.europa.eu/}
under ERC Starting Grant SECOMP (\grantnum{1}{715753}), by the Deutsche Forschungsgemeinschaft (DFG\iffull, German Research Foundation\fi)
as part of the Excellence Strategy of the German Federal and State Governments
-- EXC 2092 CASA - 390781972, by the \grantsponsor{2}{National Science Foundation}{https://www.nsf.gov/}
under grants \grantnum{2}{2048499} and \grantnum{2}{2314323}, and by the \grantsponsor{3}{European Comission}{https://commission.europa.eu/} under
grant \grantnum{3}{101070374}.
\fi

\ifcamera\else
  \pagebreak
  \onecolumn 
  \appendix
  \section*{Appendices}
\section{Calls and returns in SECOMP}\label{app:call-return}
Consider the example from \autoref{fig:example-call-return}:
during the execution of a procedure \(\mathtt{C.f}\), a call instruction to
\(\mathtt{C'.g}\) is encountered. The compartmentalized semantics first compare
the source and destination of the call.
If the call is an internal call (bottom arrow), then the state transitions to a
call-state of that compartment without generating an event. If the call is a
cross-compartment call, then the semantics perform an additional check: if the
call is allowed, then the state transitions to a call-state, but this
time generating a event recording the call (top arrow).
Otherwise, the execution must stop.

Moreover, in the case of a cross-compartment call, the stack frame that is pushed on the stack
is made read-only.
This prevents the following from happening: after a call-back to \(\mathtt{C}\),
\(\mathtt{C}\) could try using a pointer to the first stack frame to change an
argument; then, after returning to \(\mathtt{C'}\), the content of the arguments
\(\mathtt{C'}\) needs could have changed, which shouldn't be allowed.

From this call-state, there are two possibilities:
If \(\mathtt{g}\) is not a system call, the call-state transitions silently to a standard state of
\(\mathtt{C'}\) whose execution continues until reaching a return instruction,
upon which the standard state silently transitions to a return-state preparing
to return to \(\mathtt{C}\).
If \(\mathtt{g}\) is a system call, then the call-state directly transitions to a
return-state preparing to return to \(\mathtt{C}\), according to the semantics
of the system call.
Now, from the return-state, and similarly to the transition to the call-state,
we must again check whether the return is allowed, and transition to the state
of \(\mathtt{C}\) if it is the case.
The transition also restores the write access to the stack frame.
Note that in the example above, it is possible for \(\mathtt{C'}\) to return to
\(\mathtt{C}\) from a different procedure \(\mathtt{g'}\).
This may happen when, internally, \(\mathtt{g}\) performs a tail call to
\(\mathtt{g'}\).

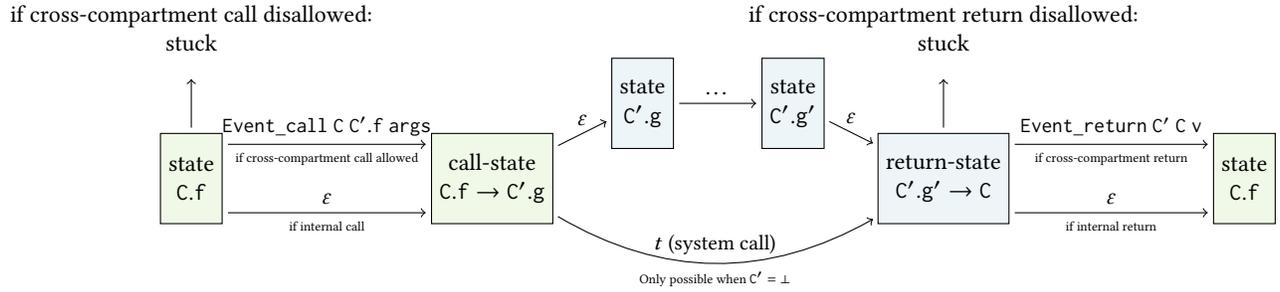
\begin{figure*}[h!]
  \definecolor{comp1}{HTML}{B2DF8A}
  \definecolor{comp2}{HTML}{A6CEE3}
  \centering
  \begin{tikzpicture}
    \node[rectangle, draw, fill=comp1!20, align=center, minimum height=1.2cm] (state) at (0,0) {state \\ \(\mathtt{C.f}\)};
    \node[rectangle, draw, fill=comp1!20, align=center, minimum height=1.2cm] (callstate) at (4, 0) {call-state \\ \(\mathtt{C.f}\rightarrow\mathtt{C'.g}\)};
    \node[rectangle, draw, fill=comp2!20, align=center, minimum height=1.2cm] (state1) at (6, 1) {state \\ \(\mathtt{C'.g}\)};
    \node[rectangle, draw, fill=comp2!20, align=center, minimum height=1.2cm] (state2) at (8, 1) {state \\ \(\mathtt{C'.g'}\)};
    \node[rectangle, draw, fill=comp2!20, align=center, minimum height=1.2cm] (returnstate) at (10, 0) {return-state \\ \(\mathtt{C'.g'}\rightarrow\mathtt{C}\)};
    \node[rectangle, draw, fill=comp1!20, align=center, minimum height=1.2cm] (state3) at (14, 0) {state \\ \(\mathtt{C.f}\)};

    \node[align=center, minimum height=1.2cm] (stuck1) at (0,2) {if cross-compartment call disallowed:\\ stuck};
    \node[align=center, minimum height=1.2cm] (stuck2) at (10,2) {if cross-compartment return disallowed:\\ stuck};

    \path (state.east) -- (state.north east) coordinate[pos=0.75] (state_ne);
    \path (callstate.west) -- (callstate.north west) coordinate[pos=0.75] (callstate_nw);
    \draw[->,shorten >= 2pt, shorten <= 2pt] (state_ne) to node [below] {\tiny if cross-compartment call allowed} (callstate_nw);
    \path[shorten >= 2pt, shorten <= 2pt] (state_ne) to node [above] {\small \(\mathtt{Event\_call~C~C'.f~args}\)} (callstate_nw);
    \path (state.east) -- (state.south east) coordinate[pos=0.75] (state_se);
    \path (callstate.west) -- (callstate.south west) coordinate[pos=0.75] (callstate_sw);
    \draw[->,shorten >= 2pt, shorten <= 2pt] (state_se) to node [below] {\tiny if internal call} (callstate_sw);
    \path[shorten >= 2pt, shorten <= 2pt] (state_se) to node [above] {\(\varepsilon\)} (callstate_sw);
    \draw[->,shorten >= 2pt, shorten <= 2pt] (state.north) to node {} (stuck1.south);

    \path (returnstate.east) -- (returnstate.north east) coordinate[pos=0.75] (returnstate_ne);
    \path (state3.west) -- (state3.north west) coordinate[pos=0.75] (state3_nw);
    \draw[->,shorten >= 2pt, shorten <= 2pt] (returnstate_ne) to node [below] {\tiny if cross-compartment return} (state3_nw);
    \path[shorten >= 2pt, shorten <= 2pt] (returnstate_ne) to node [above] {\small \(\mathtt{Event\_return~C'~C~v}\)} (state3_nw);
    \path (returnstate.east) -- (returnstate.south east) coordinate[pos=0.75] (returnstate_se);
    \path (state3.west) -- (state3.south west) coordinate[pos=0.75] (state3_sw);
    \draw[->,shorten >= 2pt, shorten <= 2pt] (returnstate_se) to node [below] {\tiny if internal return} (state3_sw);
    \path[shorten >= 2pt, shorten <= 2pt] (returnstate_se) to node [above] {\(\varepsilon\)} (state3_sw);
    \draw[->,shorten >= 2pt, shorten <= 2pt] (returnstate.north) to node {} (stuck2.south);

    \draw[->,shorten >= 2pt, shorten <= 2pt] (callstate) to node [above] {\(\varepsilon\)} (state1);
    \draw[->,shorten >= 2pt, shorten <= 2pt] (state1) to node [above] {\dots} (state2);
    \draw[->,shorten >= 2pt, shorten <= 2pt] (state2) to node [above] {\(\varepsilon\)} (returnstate);

    \draw[->,shorten >= 2pt, shorten <= 2pt,bend right] (callstate) to node [above] {\small \(t\) (system call)} (returnstate);
    \path[->,shorten >= 2pt, shorten <= 2pt,bend right] (callstate) to node [below] {\tiny Only possible when \(\mathtt{C'} = \bot\)} (returnstate);
  \end{tikzpicture}
  \caption{Call from compartment \(\mathtt{C}\) (light
    green states) to compartment \(\mathtt{C'}\) (light blue states), and the
    corresponding return.}\label{fig:example-call-return}
\end{figure*}

\section{Background: memory injections}
\emph{Memory injections}~\cite{LeroyB08} are a technical tool introduced by CompCert in order to
deal with complex program transformations that can change the memory layout of the program.
A memory injection is a partial function from block identifiers to pairs of
block identifiers and offsets.
%
Given such a memory injection \(j\), \(j\) injects values into other values in the following sense:
\(j\) injects pointer \((b, o)\) into \((b', o')\) if \(j~b = (b', z)\) and \(o' = o + z\);
\(j\) injects undefined values into any value; and
\(j\) injects non-pointer values into themselves.
Then, one can lift this relation to memories: \(j\) injects \(\mathtt{M}\) into \(\mathtt{M'}\),
written \(\mathtt{M} \mapsto_{j} \mathtt{M'}\) if,
whenever \(j~b = (b', z)\) then \(\mathtt{M}(b, o) = v \implies \mathtt{M'}(b', o + z) = v'\) with
\(j\) injecting \(v\) into \(v'\).
Memory injections are at the core of CompCert and satisfy a number of useful properties, in particular
commutation with memory operations.

\section{Axiomatization of system calls}

The formal definition of the CompCert axiom (extended with compartments)
we use in back-translation is the following:
\begin{axiom}[Commutativity of system calls and memory injections]
  Given two environments \(\mathtt{ge_{1}}\) and \(\mathtt{ge_{2}}\),
  two memories \(\mathtt{M_{1}}\) and \(\mathtt{M_{2}}\) and a memory injection
  \(j\) that relates \(\mathtt{M_{1}}\) and \(\mathtt{M_{2}}\)
  (written \(\mathtt{M_{1}} \mapsto_{j} \mathtt{M_{2}}\)) and that relates the symbols of
  the current compartment in
  the environments \(\mathtt{ge_{1}}\) and \(\mathtt{ge_{2}}\), then
  if a system call \(\mathtt{ef}\) with arguments \(\mathtt{args}\) produces
  a trace \(t\) in \(\mathtt{ge_{1}}\) and \(\mathtt{M_{1}}\), resulting in return value \(\mathtt{res}\)
  and new memory \(\mathtt{M_{1}'}\),
  \[
    \mathtt{ge_{1}}, \mathtt{M_{1}} \vdash \mathtt{ef}~\mathtt{args} \xrightarrow{t} (\mathtt{res}, \mathtt{M_{1}'})
  \]
  then:
  \begin{align*}
    \exists~j'~\mathtt{res'}~\mathtt{M_{2}'}.~
    \mathtt{ge_{2}}, \mathtt{M_{2}} \vdash \mathtt{ef}~\mathtt{args} \xrightarrow{t} (\mathtt{res'}, \mathtt{M_{2}'})
  \end{align*}
  with \(\mathtt{M_{1}'} \mapsto_{j'} \mathtt{M_{2}'}\).
\end{axiom}

\section{Details of the back-translation function}

In this appendix, we describe the back-translation function in full details.

Recall the 3 kinds of informative events:
\begin{enumerate}[leftmargin=15pt,nosep]
  \item $\mathtt{Icall~f~t~g~\overline{v}~sg}~\Delta$ represents a cross-compartment call,
  where $\mathtt{f}$ is the name of the caller,
  $\mathtt{t}$ is the trace event produced by the call,
  $\mathtt{g}$ is the name of the callee,
  $\mathtt{\overline{v}}$ are the arguments,
  $\mathtt{sg}$ is the signature of the callee,
  and $\Delta$ are the memory deltas.
  \item $\mathtt{Ireturn~f~t~{v}}~\Delta$ represents a cross-compartment return,
  where $\mathtt{f}$ is the current procedure name,
  $\mathtt{t}$ is the trace event produced by the return,
  $\mathtt{v}$ is the return value,
  and $\Delta$ are 
  memory deltas.
  \item $\mathtt{Isys~f~t~ef~\overline{v}}~\Delta$ represents a system call,
  which is similar to the call case except that it requires the
  system call descriptor $\mathtt{ef}$
  instead of $\mathtt{g}$ and $\mathtt{sg}$.
\end{enumerate}
where \(\Delta\) represents a list of memory deltas.
Formally, the memory deltas are defined as follows:
\[
  \begin{array}{ll}
    \delta \mathrel{:=}&|~\mathtt{delta\_storev~ch~p~v~C} \\
                       &|~\mathtt{delta\_store~ch~b~o~v~C} \\
                     &|~\mathtt{delta\_bytes~b~o~\overline{v}~C} \\
                     &|~\mathtt{delta\_alloc~C~l~h} \\
                      &|~\mathtt{delta\_free~b~l~h~C}
  \end{array}
\]
where stores, storevs and bytes are three different kind of stores, while allocs
and frees corresponds to the same memory operations.

The back-translation function is defined piecewise.
First, the back-translation of memory deltas in compartment \(\mathtt{C}\) \(\mathtt{bt^{C}_{\delta}}\) is defined as follows:
\[
\begin{array}{ll}
  \mathtt{bt^{C}_{\delta}}\left(\mathtt{delta\_storev~ch~p~v~C}\right) &= \mathtt{if~symbol\_of\left(p\right) = id \wedge public^{C}\left(id\right)
                                                                                       \wedge not\_pointer\left(v\right)}\\
                                                                       & \mathtt{\quad\quad then~p := v~else~skip} \\
  \mathtt{bt^{C}_{\delta}}\left(\mathtt{delta\_store~ch~b~o~v~C}\right) &= \mathtt{if~symbol\_of\left(b\right) = id \wedge public^{C}\left(id\right) \wedge not\_pointer\left(v\right)}\\
                                                                       & \mathtt{\quad\quad then~id[o] := v~else~skip}\\
  \mathtt{bt^{C}_{\delta}}\left(\mathtt{delta\_bytes~b~o~\overline{v}~C}\right) &= \mathtt{if~symbol\_of\left(b\right) = id \wedge public^{C}\left(id\right)
                                                                                               \wedge not\_pointer\left(\overline{v}\right)}\\
                                                                       & \mathtt{\quad\quad then~id := \overline{v}~else~skip}\\
  \mathtt{bt^{C}_{\delta}}\left(\mathtt{delta\_alloc~C~l~h}\right) &= \mathtt{skip} \\
  \mathtt{bt^{C}_{\delta}}\left(\mathtt{delta\_free~b~l~h~C}\right) &= \mathtt{skip}
\end{array}
\]
where \(\mathtt{symbol\_of}\) returns the identifier of its argument, if it exists,
\(\mathtt{public^{C}\left(id\right)}\) is \(\mathtt{true}\) if and only if \(\mathtt{id}\)
is an identifier corresponding to a public global variable of \(\mathtt{C}\), and \(\mathtt{not\_pointer\left(v\right)}\)
is \(\mathtt{true}\) if and only if \(\mathtt{v}\) is not a pointer.

We define the back-translation of a list of memory deltas in \(\mathtt{C}\), \(\mathtt{bt^{C}_{\mathnormal{\Delta}}}\) as follows:
\begin{align*}
  \mathtt{bt^{C}_{\mathnormal{\Delta}}\left([]\right)} &= \mathtt{skip} \\
  \mathtt{bt^{C}_{\mathnormal{\Delta}}\left(\delta~::~\mathnormal{\Delta}\right)} &= \mathtt{bt^{C}_{\delta}\left(\delta\right)~;;~bt^{C}_{\mathnormal{\Delta}}\left(\mathnormal{\Delta}\right)}
\end{align*}

Note that we discard the stores to public global variables that are not scalar; this is not an issue,
as our well-formedness conditions ensure that the last store made to a global variable before issuing
an event is necessarily that of a scalar.

We define the back-translation of events in compartment \(\mathtt{C}\), \(\mathtt{bt^{C}_{e}}\) as follows:
\begin{align*}
  \mathtt{bt^{C}_{e}}\left(\mathtt{Icall~f~t~g~\overline{v}~sg}~\Delta\right) &=
    \mathtt{bt^{C}_{\mathnormal{\Delta}}\left(\mathnormal{\Delta}\right)~;;~g\left(\overline{v}\right)}\\
  \mathtt{bt^{C}_{e}}\left(\mathtt{Ireturn~f~t~{v}}~\Delta\right) &=
    \mathtt{bt^{C}_{\mathnormal{\Delta}}\left(\mathnormal{\Delta}\right)~;;~return~v}\\
  \mathtt{bt^{C}_{e}}\left(\mathtt{Isys~f~t~ef~\overline{v}}~\Delta\right) &=
    \mathtt{bt^{C}_{\mathnormal{\Delta}}\left(\mathnormal{\Delta}\right)~;;~ef\left(\overline{v}\right)}\\
\end{align*}

We can finally define, for a given compartment \(\mathtt{C}\), the back-translation of a list of events from that compartment,
\(\mathtt{bt^C}\) in the following manner:
\begin{align*}
  \mathtt{bt^{C}_{n}}\left(\mathtt{[]}\right) &= \mathtt{return}\\
  \mathtt{bt^{C}_{n}}\left(\mathtt{e :: ls}\right) &= \mathtt{if~C.ctr = n}\\
                                              &\quad\quad\mathtt{~then~C.ctr++~;;~bt^{C}_{e}}\left(\mathtt{e}\right)~\\
                                              &\quad\quad~\mathtt{else}~\mathtt{bt^{C}_{n+1}\left(\mathtt{ls}\right)}\\
  \mathtt{bt^{C}} &= \mathtt{bt^{C}_{0}}
\end{align*}
where \(\mathtt{C.ctr}\) is a counter local to compartment \(\mathtt{C}\).

Then, each back-translated procedure is defined by wrapping this in a \(\mathtt{while~true}\) loop.

\section{Blame Proof Details}
\label{app:blame}

\rb{Intro here}

The simulation invariants again combine \emph{strong} and \emph{weak}
relations, both parameterized by a \emph{partial memory injection} that
is built throughout the runs.
\rb{I'm not convinced that talking about this in strong-weak terms is
particularly clear; it's not the same as in recomposition.}
The strong relation holds when the shared context part is in charge of
lockstep execution in both programs, and relates the various \emph{state
components} (memories, continuations, local environments, call arguments
and return values) according to the injection.
The weak relation, in turn, holds when each program part is in charge of
execution in its own run. During this time, it relates the
\emph{context-side parts} of both memories and continuations, but
ignores differences between the two running program sides.

\rb{This feels too detailed, definitely move to technical section}
In both cases, a partial memory injection records a mapping between the
memory blocks of both executions, but disregarding most of the contents
of the program side, only recording the correspondence between the
\emph{public symbols} of both programs.
As noted, this partial injection does not remain constant across the
runs, but needs to be updated as relevant blocks are newly allocated.

Let $\mathcal{R}_j$ denote the combined strong-weak simulation relation
under injection $j$.
\rb{\ldots}

At the most basic level, we employ three properties based on individual
\clight steps.

\begin{enumerate}\small
  \item \(\forall~j~s_1~s_1'~s_2~s_2'~a.~\mathcal{R}_j~s_1~s_2 \implies s_1 \in K \implies
  s_1 \xrightarrow{a} s_{1}' \implies s_2 \xrightarrow{a} s_{2}' \implies
  \exists j'.~\mathcal{R}_{j'}~s_1'~s_2' \)
  \item \(\forall~j~s_1~s_1'~s_2~s_2'~a.~\mathcal{R}_j~s_1~s_2 \implies s_1 \notin K \implies
  s_1 \xrightarrow{a} s_{1}' \implies s_2 \xrightarrow{a} s_{2}' \implies
  \exists j'.~\mathcal{R}_{j'}~s_1'~s_2' \)
  \item \(\forall~j~s_1~s_1'~s_2.~\mathcal{R}_j~s_1~s_2 \implies s_1 \notin K \implies
  s_1 \xrightarrow{\epsilon} s_{1}' \implies
  \exists j'.~\mathcal{R}_{j'}~s_1'~s_2 \)
\end{enumerate}

Properties (1) and (2) state that the relation $\mathcal{R}$ is
preserved when \emph{each} of the two programs takes a \emph{single
step}, each producing the \emph{same trace}---respectively, from the
context side or from the program side.
Property (3) further shows that $\mathcal{R}$ is preserved when
just \emph{one} of the programs takes a single \emph{silent step} from
the \emph{program side} while the other stays put.
\rb{Could give a bit of intuition about why these properties hold at
the basic level. The one-by-one descriptions could be more readable
but are of course redundant.}

\begin{enumerate}
\item $\mathcal{R}$ is preserved when \emph{each} of the two programs takes a
\emph{single step} from the \emph{context side}, each producing the
\emph{same trace}.
\item $\mathcal{R}$ is preserved when each of the two programs takes a
single step from the \emph{program side}, each producing the same
trace.
\item $\mathcal{R}$ is preserved when \emph{one} of the programs
takes a single \emph{silent step} from the \emph{program side} while the
other stays put.
\end{enumerate}

\rb{...}

\begin{enumerate}\small
  \item \(\forall~j~s_1~s_1'~s_1''~s_2~s_2'~s_2''~a.~\mathcal{R}_j~s_1~s_2 \implies s_1 \in K \implies
  s_1 \xrightarrow{\epsilon}^{*} s_{1}' \implies s_1' \xrightarrow{a} s_{1}'' \implies
  s_2 \xrightarrow{\epsilon}^{*} s_{2}' \implies s_2' \xrightarrow{a} s_{2}'' \implies
  \exists j'. \mathcal{R}_{j'}~s_1''~s_2'' \)
  \item \(\forall~j~s_1~s_1'~s_1''~s_2~s_2'~s_2''~a.~\mathcal{R}_j~s_1~s_2 \implies s_1 \notin K \implies
  s_1 \xrightarrow{\epsilon}^{*} s_{1}' \implies s_1' \xrightarrow{a} s_{1}'' \implies
  s_2 \xrightarrow{\epsilon}^{*} s_{2}' \implies s_2' \xrightarrow{a} s_{2}'' \implies
  \exists j'. \mathcal{R}_{j'}~s_1''~s_2'' \)
  \item \(\forall~j~s_1~s_1''~s_2~s_2''~t~t_1~t_2.~\mathcal{R}_j~s_1~s_2 \implies
  s_1 \xrightarrow{t \cdot t_1}^{*} s_{1}'' \implies
  s_2 \xrightarrow{t \cdot t_2}^{*} s_{2}'' \implies
  \exists j'~s_1'~s_2'.~s_1 \xrightarrow{t}^{*} s_{1}' \wedge s_{1}' \xrightarrow{t_1}^{*} s_{1}'' \wedge
  s_2 \xrightarrow{t}^{*} s_{2}' \wedge s_{2}' \xrightarrow{t_2}^{*} s_{2}'' \wedge
  \mathcal{R}_{j'}~s_1'~s_2' \)
\end{enumerate}

\begin{enumerate}
\item $\mathcal{R}$ is preserved when \emph{each} of the two programs
takes an arbitrary sequence of \emph{silent steps} from the
\emph{context side}, followed by a pair of \emph{synchronous steps}
producing the \emph{same event}.
\emph{same trace}.
\item $\mathcal{R}$ is preserved when each of the two programs
takes an arbitrary sequence of silent steps from the
\emph{program side}, followed by a pair of synchronous steps
producing the same event.
\emph{same trace}.
\item $\mathcal{R}$ is preserved after any \emph{shared trace prefix}
produced by a pair of executions of the two programs.
\end{enumerate}

\rb{\ldots}

\begin{enumerate}\small
  \item \(\forall~j~s_1~s_1'~s_2~s_2'~t~t'.~\mathcal{R}_j~s_1~s_2 \implies
  s_1 \xrightarrow{t \cdot t'}^{*} s_{1}' \implies
  s_2 \xrightarrow{t}^{*} s_{2}' \implies
  s_1' \textrm{final}~ \implies
  s_2' \not\xrightarrow{} \implies s_2' \in K \implies
  s_2' \textrm{final} \)
  \item \(\forall~j~s_1~s_1'~s_2~s_2'~t~a~t'.~\mathcal{R}_j~s_1~s_2 \implies
  s_1 \xrightarrow{t \cdot a :: t'}^{*} s_{1}' \implies
  s_2 \xrightarrow{t}^{*} s_{2}' \implies
  s_2' \not\xrightarrow{} \implies
  s_2' \notin K \)
\end{enumerate}

\rb{Technical details moved from overview}
All our definitions\ch{What definitions? (bad flow) Actually, no clue why are
  you telling me any of the remaining stuff in this paragraph either.  So at the
  very least you should explain the motivation for all this and how it relates
  to what you wrote so far.}\rb{Agreed! moved down for now}
are specialized to work with our compartmentalized
version of \clight.
\rb{This is currently somewhat out of order: mostly described in section
3 but also anticipated for the ASM case in the introduction to
recomposition -- talking about them here breaks the flow somewhat}
Program states in \clight always include the current memory and
continuation (which abstracts the call stack), with additional
information depending on the type of state: standard states add the
executing procedure, the language statement being evaluated and local
variable environments; call-states replace those with information
about the callee and its arguments; and return-states with information
about the return value.
\fi

\bibliographystyle{plainurl}
\balance
\bibliography{mp,safe}

%

\end{document}